# On the $H_2$ abundance and ortho-to-para ratio in Titan's troposphere

Bruno Bézard[*], Sandrine Vinatier

LESIA, Observatoire de Paris, Université PSL, CNRS, Sorbonne Université, Université Paris-Diderot, Sorbonne Paris Cité, 5 place Jules Janssen, 92195 Meudon, France

[*] Corresponding author.
   *E-mail address:* Bruno.Bezard@obspm.fr

# ABSTRACT

We have analyzed spectra recorded between 50 and 650 cm$^{-1}$ by the Composite Infrared Spectrometer (CIRS) aboard the Cassini spacecraft at low and high emission angles to determine simultaneously the H$_2$ mole fraction and ortho-to-para ratio in Titan's troposphere. We used constraints from limb spectra between 50 and 900 cm$^{-1}$ and from in situ measurements by the Huygens probe to characterize the temperature, haze and gaseous absorber profiles. We confirm that the N$_2$-CH$_4$ collision-induced absorption (CIA) coefficients used up to now need to be increased by about 52% at temperatures of 70-85 K. We find that the N$_2$-N$_2$ CIA coefficients are also too low in the N$_2$ band far wing, beyond 110 cm$^{-1}$, in agreement with recent quantum mechanical calculations. We derived a H$_2$ mole fraction equal to (0.88 ± 0.13) × 10$^{-3}$, which pertains to the ~1-34 km altitude range probed by the $S_0$(0) and $S_0$(1) lines. This result agrees with a previous determination based only on the H$_2$-N$_2$ dimer transition in the $S_0$(0) line, and with the in situ measurement by the Gas Chromatograph Mass Spectrometer (GCMS) aboard Huygens. It is 3-4 times smaller than the value measured in situ by the Ion Neutral Mass Spectrometer (INMS) of Cassini at 1000-1100 km. The H$_2$ para fraction is close to equilibrium in the 20-km region. CIRS spectra can be fitted assuming ortho-to-para (o-p) H$_2$ thermodynamical equilibrium at all levels or a constant para fraction in the range 0.49-0.53. We have investigated different mechanisms that may operate in Titan's atmosphere to equilibrate the H$_2$ o-p ratio and we have developed a one-dimensional model that solves the continuity equation in presence of such conversion mechanisms. We conclude that exchange with H atoms in the gas phase or magnetic interaction of H$_2$ in a physisorbed state on the surface of aerosols are too slow compared with atmospheric mixing to play a significant role. On the other hand, magnetic interaction of H$_2$ with CH$_4$, and to a lesser extent N$_2$, can operate on a timescale similar to the vertical mixing time in the troposphere. This process is thus likely responsible for the o-p equilibration of $H_2$ in the mid-troposphere implied by CIRS measurements. The model can reproduce the inferred o-p ratio in the 20-km region, assuming low atmospheric mixing in the troposphere down to 15-20 km and conversion rates with $CH_4$ or $N_2$ slightly larger than obtained from an extrapolation of natural ortho-para conversion rate measured in gaseous hydrogen.

# 1. Introduction

Molecular hydrogen ($H_2$) is formed in the upper atmosphere of Titan from the photodissociation of methane ($CH_4$) and, to a lesser extent, of other hydrocarbons in the stratosphere. This occurs either directly, or indirectly from reaction of H with various radicals (Lavvas et al. 2008, Krasnopolsky 2009, 2012, 2014, Vuitton et al. 2019). $H_2$ is chemically inert and, in steady state, production must be balanced by escape from the atmosphere, assuming no loss at the surface (Strobel 2010).

Measurements with the Cassini Composite Infrared Spectrometer (CIRS) aboard Cassini yielded a globally averaged $H_2$ mole fraction of $(9.6 \pm 2.4) \times 10^{-4}$ in Titan's upper troposphere (Courtin et al. 2012). This value agrees with previous determinations from Voyager infrared measurements by the infrared instrument (IRIS) (Courtin et al. 1995, Samuelson et al. 1997). It is also fully consistent with the in situ measurement by the Gas Chromatograph Mass Spectrometer (GCMS) on the Huygens probe: $(1.01 \pm 0.16) \times 10^{-3}$ in the atmosphere below 140 km (Niemann et al. 2010). In situ measurements by the Ion Neutral Mass Spectrometer (INMS) aboard Cassini recorded during 15 flybys of Titan allowed Cui et al. (2009) to derive globally-averaged vertical profiles of $N_2$, $CH_4$ and $H_2$ densities between 981 and 1151 km. The corresponding $H_2$ mixing ratio increases from $(3.30 \pm 0.01) \times 10^{-3}$ to $(4.28 \pm 0.01) \times 10^{-3}$ over this altitude range. Combined with a diffusion model, this profile implies a $H_2$ escape flux of $\sim 1.1 \times 10^{10}$ cm$^{-2}$ s$^{-1}$ (referenced to Titan's surface) (Cui et al. 2009), which is essentially equal to Hunten's limiting flux (Strobel 2012). This flux, together with Courtin et al's (2012) $H_2$ globally-averaged mole fraction, yields a $H_2$ lifetime in Titan's atmosphere ($\tau_{esc}$, equal to the ratio of the column density to the flux) as long as $\sim 7 \times 10^5$ years ($2 \times 10^{13}$ s), much longer than the atmospheric mixing time. Therefore, given its chemical stability, we expect $H_2$ to be uniformly mixed throughout the atmosphere below the homopause, which is located around 800 km. From detailed model calculations incorporating eddy, molecular and thermal diffusion along with a simplified chemical model, Strobel (2010, 2012) concluded that the tropospheric (from CIRS) and thermospheric (from INMS) values are incompatible by a factor of 2-3, in the lack of $H_2$ downward flux at the surface. Similarly, the detailed photochemical calculations of Vuitton et al. (2019) show that a model that reproduces Courtin et al.'s $H_2$ tropospheric mole fraction has $H_2$ mole fractions about half lower than the INMS measurements (see their Fig. 63). In contrast, Krasnopolsky's (2012, 2014) model is able to reproduce both the observed GCMS value ($1 \times 10^{-3}$) below 140 km and the INMS data at 1000-1200 km thanks to its eddy mixing profile, which exhibits a second homopause near 550 km.

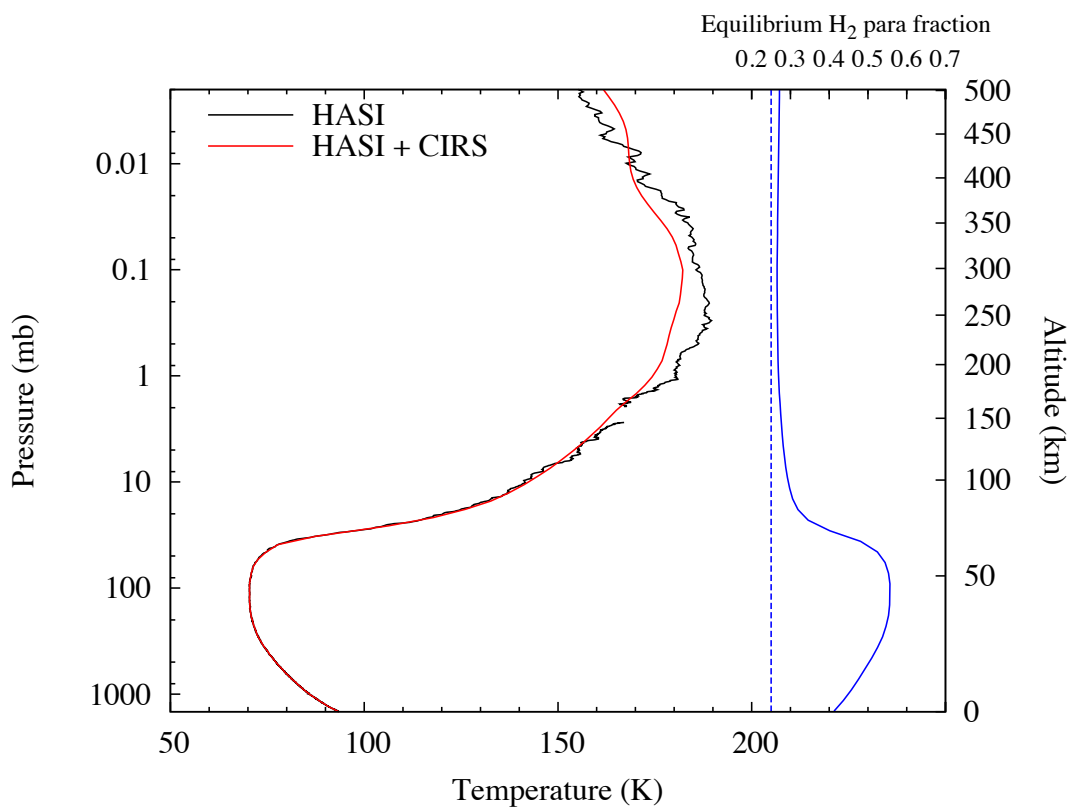

**Fig. 1.** *Left.* Temperature profiles from Huygens/HASI (Fulchignoni et al. 2005) (black) and with constraints from Cassini/CIRS measurements above ~100 km (red). *Right.* Vertical profile of the equilibrium para-fraction of $H_2$ in Titan's atmosphere (solid blue). The dashed line represents the high-temperature limit (0.25).

Molecular hydrogen exists in two forms: a singlet state with an anti-parallel configuration of the two proton spins (para-hydrogen) and a triplet state with the proton spins aligned parallel (ortho-hydrogen). Due to the antisymmetry of the total $H_2$ wave function under nuclear exchange, para-$H_2$ is associated with even rotational numbers ($J$) and ortho-$H_2$ is associated with odd $J$. At thermodynamical equilibrium, the $H_2$ para fraction is given by

$$f_{para}^{eq} = \frac{z_{para}}{z_{para} + 3\, z_{ortho}}, \tag{1}$$

where $z_{para}$ and $z_{ortho}$ are the rotational partition functions of para-$H_2$ (even $J$) and ortho-$H_2$ (odd $J$). Therefore, at high temperatures (typically $T > 200$ K), $f_{para}^{eq} = 0.25$, while $f_{para}^{eq} = 1$ at $T = 0$. Because transitions with $\Delta J = 1$ are forbidden, conversion between ortho and para states is inhibited and thermodynamical equilibration between the two species is very slow in the lack of interaction with other substances.

The CIRS-based determination of Courtin et al. (2012) is based on $N_2$-$H_2$ dimer transitions occurring in the region of the $S_0(0)$ rotational transition of $H_2$ near 355 cm$^{-1}$. This transition, connecting even rotational numbers, is due to para-$H_2$. To retrieve the $H_2$ mole fraction, it is thus necessary to make some assumption on the ortho-to-para ratio and Courtin et al. assumed local thermodynamic equilibrium between the two forms. Determination of the ortho-$H_2$ abundance using Cassini/CIRS measurements in the $S_0(1)$ rotational transition near 590 cm$^{-1}$ is difficult. This region is located at the long-frequency edge of Focal Plane 1 (FP1) and at the short-wavelength edge of Focal Plane 3 (FP3), where the sensitivity of the instrument is poor. In addition, the flux from Titan's atmosphere near 590 cm$^{-1}$ is weak and mixes tropospheric emission sensitive to $H_2$ absorption and stratospheric haze emission, which thus need to be disentangled.

The $H_2$ mole fraction is equal to the para-$H_2$ mole fraction $q_{para}$ (constrained by the $S_0(0)$ absorption) divided by the para fraction in $H_2$ ($f_{para}$). $H_2$ is formed from photochemistry in the thermosphere (around 800 km) and stratosphere (around 200-300 km) (Strobel 2010, Vuitton et al. 2019) where the temperature is relatively high, 150-170 K, leading to $f_{para}$ at the formation of 0.27-0.28. However, equilibration between ortho and para $H_2$ may subsequently occur through various mechanisms. The equilibrium para fraction varies in Titan's atmosphere at low latitudes between 0.26 near the stratopause (185 K) and 0.56 at the temperature minimum (70 K) (Fig. 1). Its value in the region probed by the $S_0(0)$ and $S_0(1)$ lines, mostly in the 78-85 K temperature range (Courtin et al. 2012), is ~0.50-0.46. For ortho-para equilibrium to be achieved, the conversion time between the two states ($\tau_{conv}$) has to be smaller than the dynamical mixing time ($\tau_{dyn}$) shown in Fig. 11. Based on one-dimensional photochemical modelling (Lavvas et al. 2008, Vuitton et al. 2019), $\tau_{dyn}$, which we define as $H^2 / K$ where $H$ is the pressure scale height and $K$ the eddy mixing coefficient, varies between $10^6$ s at 400-700 km and a few $10^9$ s at 40-80 km (Fig. 11). On the other hand, ortho-to-para conversion times can be very long. Departure from local thermodynamical equilibrium has been inferred in the upper tropospheres of the giant planets from analyses of the collision-induced hydrogen absorption in Voyager infrared spectra (Conrath and Gierasch 1984, Conrath et al. 1998) and in their stratospheres from observations of $H_2$ quadrupole lines with the Infrared Space Observatory (Fouchet et al. 2003). Conrath et al. (1998) estimated the conversion times from these investigations to be 3-5 × $10^9$ s in the 200-mbar region. If the ortho-para conversion times on Titan were similar or larger, then, most likely, $\tau_{dyn} < \tau_{conv} < \tau_{esc}$. In such a case, we expect $f_{para}$ to be essentially uniform in the atmosphere and fixed by equilibration in regions where $\tau_{conv} / \tau_{dyn}$ is at minimum. The reason is that, in such regions, $H_2$ in an air parcel can more efficiently undergo (partial) conversion before being transported away, a process equivalent to "chemical quenching". In a situation where conversion is more strongly inhibited, $\tau_{conv} > \tau_{esc}$, then $f_{para}$ in the atmosphere would retain its value at formation (0.27-0.28).

It is thus conceivable that $H_2$ is not at thermodynamical equilibrium in Titan's atmosphere, in particular in the formation region of the $S_0(0)$ line, implying that the CIRS-derived $H_2$ mole fraction may be in error. For example, if $f_{para}$ is close to 0.28, a value representative of equilibrium in the upper stratosphere, rather than 0.50, typical of the mid-troposphere where the $S_0(0)$ line is formed, then the $H_2$ mole fraction could be almost twice larger than derived by Courtin et al. (2012), which would solve most of the discrepancy between the CIRS and INMS determinations. This was the motivation of the present study. It is also worth noting that Courtin et al. found a $H_2$ mole fraction above 50°N in excess of 30-70% relatively to the globally-averaged value. They proposed that this enhancement results from downwelling in the winter season in presence of a vertical gradient of the $H_2$ mole fraction in the stratosphere. However, this enhancement, derived from the $S_0(0)$ line, actually corresponds to an enhancement of the *para* $H_2$ mole fraction. The existence and amplitude of the $H_2$ enhancement thus critically depends on a possible variation of the ortho-to-para ratio at high northern latitudes which could arise from downwelling or different equilibration conditions in the troposphere.

In an analysis of Voyager IRIS spectra in the range 200-600 cm$^{-1}$, Courtin et al. (1995) concluded that there was no evidence for a departure from ortho-para $H_2$ equilibrium. They found that the high-temperature limit of $f_{para}$ (0.25) yielded a slightly worse agreement with the data but their investigation shows that the determination of the ortho-$H_2$ abundance using the $S_0(1)$ line near 590 cm$^{-1}$ is very inaccurate (see their Fig. 18a). From another analysis of Voyager spectra, Samuelson et al. (1997) concluded that the hydrogen para fraction ($f_{para}$) was close to equilibrium, with considerable uncertainty. They derived $f_{para}$ = 0.48 ± 0.10, which corresponds to a 1-SD (standard deviation) range for equilibrium temperature of 70-100 K that encompasses the $H_2$ line formation regions. They also considered the possibility of an ethane cloud near or just above the tropopause, showing that it would significantly increase their best fit value of $f_{para}$. In addition, both studies inferred a high supersaturation of methane in the troposphere of Titan, which was not observed by the Huygens/GCMS (Niemann et al. 2010). Therefore, we think that these analyses of Voyager IRIS do not provide strong and reliable constraints of the ortho-to-para ratio in $H_2$, given the uncertainties in their atmospheric models (methane profile and aerosol properties and distribution).

In situ measurements by the GCMS (Niemann et al. 2010), Huygens Atmospheric Science Instrument (HASI) (Fulchignoni et al. 2005) and Descent Imager/Spectral Radiometer (DISR) (Doose et al. 2016) on 14 January 2005 provided unique ground-truth information on the vertical profiles of methane, temperature and aerosol density below 140 km, at a specific location near 10°S. In addition, observations by CIRS at latitudes close to 10°S not too long after the Huygens descent can be used to complement the DISR measurements and constrain the wavelength-dependence and vertical profile of aerosol extinction (Anderson and Samuelson, 2011; Vinatier et al., 2010; Vinatier et al., 2012). This information is key to model adequately the far-infrared emission spectrum of Titan in the regions of the $H_2$ lines.

The paper is organized as follows. In Section 2, we investigate the constraints on the $H_2$ ortho-to-para ratio that can be derived from the Cassini CIRS spectra in combination with the Huygens measurements and provide a new estimate of the $H_2$ mole fraction. In Section 3, we survey the mechanisms for ortho-to-para conversion in Titan's atmosphere and model the vertical profile of $f_{para}$ for different assumptions on some key parameters. In Section 4, we discuss the results from Section 2 (retrieval of $H_2$ ortho-to-para ratio from CIRS spectra) and Section 3 (modelling of the $H_2$ para fraction vertical profile). We present our conclusions in Section 5.

## 2. Determination of the $H_2$ para fraction from Cassini CIRS measurements

*2.1. Observations*

The Cassini/CIRS instrument consists of two Michelson interferometers with three different focal planes. In the far infrared, Focal Plane 1 (FP1) covers the spectral range 10-650 cm$^{-1}$, while, in the mid-infrared, Focal Plane 3 (FP3) and Focal Plane 4 (FP4) cover respectively the 580-1100 and 1050-1500 cm$^{-1}$ portions of the spectrum (Flasar et al. 2004, Jennings et al. 2017). The spectral resolution can be adjusted between 0.5 and 15.5 cm$^{-1}$ (apodized). FP1 has a 3.9-mrad circular field of view (FOV) with a half-peak diameter of 2.5 mrad. FP3 and FP4 each incorporate a linear array of 10 pixels with 0.27-mrad FOV per pixel.

The $S_0(0)$ $H_2$ line around 355 cm$^{-1}$ is located well within the FP1 spectral range while the $S_0(1)$ line near 590 cm$^{-1}$ is at the edge of the FP1 bandpass, where the Noise Equivalent Spectral Radiance (NESR) is ~4 times larger than at 355 cm$^{-1}$. However, the $S_0(1)$ line is also marginally covered by the FP3, which has a lower NESR beyond 575 cm$^{-1}$, and we have then chosen to use both FP1 and FP3 to analyze the $H_2$ absorption in Titan's spectrum. Using FP3 also allows us to better constrain the $C_4H_2$ and $CH_3C_2H$ and abundances through their strong bands at 628 and 633 cm$^{-1}$, which is important because emission from these species is also seen in the FP1 range at 220 and 327 cm$^{-1}$ respectively, close to the $S_0(0)$ $H_2$ line.

We chose to analyze CIRS spectra at 15.5 cm$^{-1}$-resolution because the collision-induced absorptions from the $S_0(0)$ and $S_0(1)$ H$_2$ lines are intrinsically relatively broad. This allows us to ensure a better S/N ratio than with higher spectral resolution spectra. For a surface-intercepting (nadir) observation with FP1, the S/N ratio is only ~1 at 590 cm$^{-1}$ per spectrum. It is thus necessary to average a large number of FP1 spectra to reach a satisfactory S/N ratio in the region of the $S_0(1)$ line. We selected spectra recorded between 2005 and 2007 in the 20°S-20°N latitude range, i.e. close in latitude and time to the Huygens landing site. In this latitude range and over this time period, seasonal variations of temperature and gas and aerosol abundances are weak (Teanby et al. 2008, Coustenis et al. 2013, Bézard et al. 2018), so that we can use constraints from the in situ measurements made by the Huygens probe. We used two selections of nadir spectra to analyze the H$_2$ absorption, one with a low averaged emission angle (13°), and the second one with an average emission angle of 59°. This allows us to better disentangle the contribution from the collision-induced absorption in the troposphere and that from stratospheric emission, mainly due to the haze, as demonstrated by Samuelson et al. (1997). In addition, we made two limb-viewing selections, at altitudes centered at 105 and 145 km, to better constrain the main haze and the nitrile haze spectral opacities in the stratosphere. The characteristics of all these selections are listed in Table 1. Spectra were extracted from the CIRS v4.3.4 database in May 2018.

**Table 1.**

Characteristics of the CIRS selections

| Selection | Date range | Latitude range | Distance range (km) | Emission angle or altitude range | Number of averaged spectra | Mean emission angle or altitude |
|---|---|---|---|---|---|---|
| FP1 low emission angle | 2005 – 2007 | 20°S – 20°N | 40000 – 80000 | 0 – 30° | 10450 | 13° |
| FP3 low emission angle | 2005 – 2007 | 20°S – 20°N | 40000 – 80000 | 0 – 30° | 45993 | 12° |
| FP1 high emission angle | 2005 – 2007 | 20°S – 20°N | 40000 – 80000 | 55 – 70° | 141 | 59° |
| FP3 high emission angle | 2005 – 2007 | 20°S – 20°N | 40000 – 80000 | 55 – 70° | 678 | 59.5° |
| FP1 105-km limb | 2005 – 2007 | 20°S – 20°N | 10000 – 17000 | 90 -– 120 km | 68 | 105.5 km |
| FP3 105-km limb | 2005 – 2007 | 20°S – 20°N | 10000 – 17000 | 90 -– 120 km | 89 | 107.5 km |
| FP1 145-km limb | 2005 – 2007 | 20°S – 20°N | 10000 – 17000 | 130 -– 160 km | 50 | 145 km |
| FP3 145-km limb | 2005 – 2007 | 20°S – 20°N | 10000 – 17000 | 130 -– 160 km | 88 | 145 km |

The FP1 and FP3 averaged spectra with a surface-intercepting line-of-sight are displayed in Fig. 2 and those in the limb-viewing geometry are displayed in Fig. 3, along with the 1-SD error bars. In these plots, the step is 5 cm$^{-1}$, 3 times smaller than the spectral resolution. For all four observing geometries, the FP1 and FP3 spectra show consistent radiances in the region where they overlap (~575-670 cm$^{-1}$) given the error bars. As the FP3 noise level is smaller than that of the FP1 spectra at 580 cm$^{-1}$ and beyond, we built four composite spectra consisting of the FP1 radiances up to 575 cm$^{-1}$ and the FP3 radiances at and above 580 cm$^{-1}$.

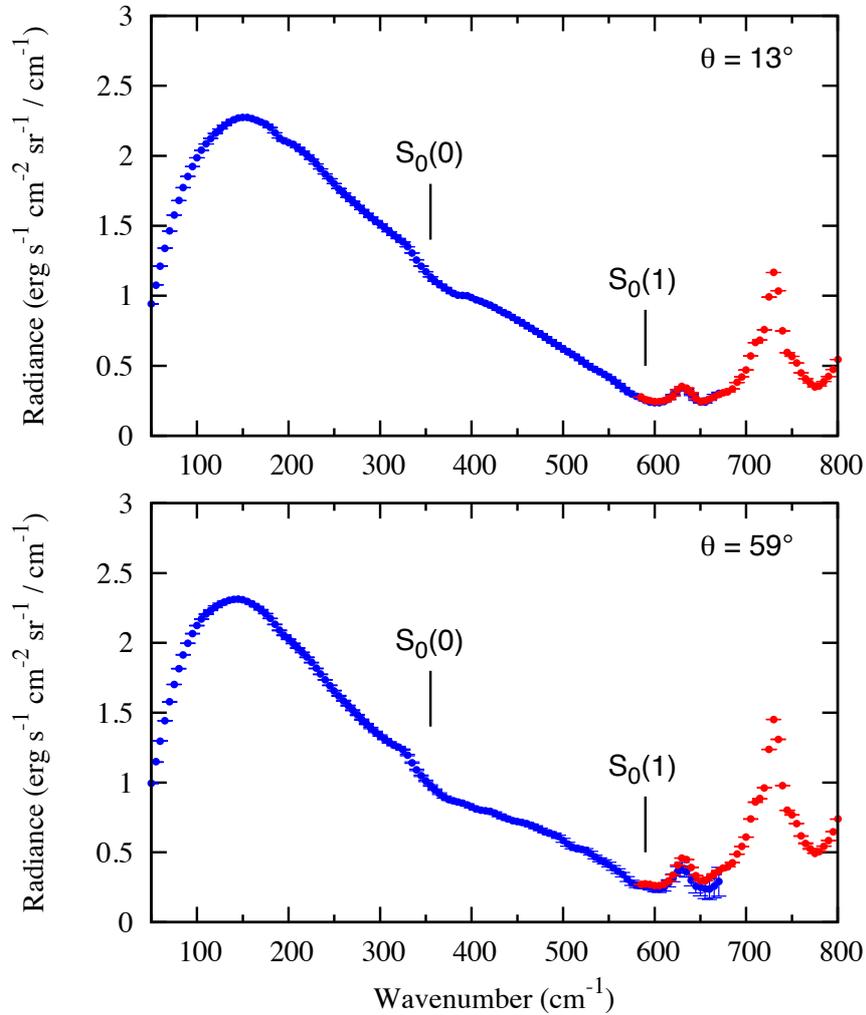

**Fig. 2.** Averages of Cassini/CIRS spectra recorded in 2005-2007 in the latitude range [20°S, 20°N] with FP1 (blue) and FP3 (red). The spectral selections in the upper panel correspond to a mean surface emission angle of 13° (airmass of 1.02) and those in the lower panel to an emission angle of 59° (airmass of 1.96). The error bars correspond to the 1-SD Noise Equivalent Spectral Radiance. At most wavenumbers, they are smaller than the dot size.

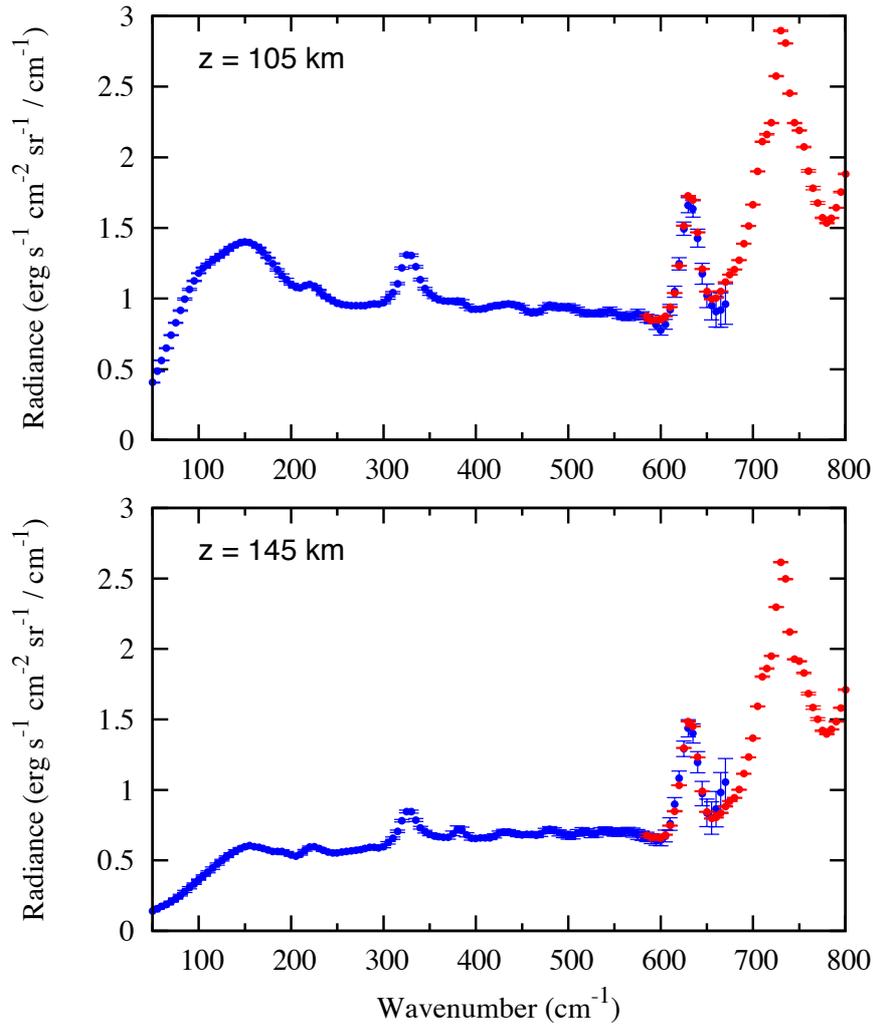

**Fig. 3.** Averages of Cassini/CIRS limb-viewing spectra recorded in 2005-2007 in the latitude range [20°S, 20°N] with FP1 (blue) and FP3 (red). The spectral selections in the upper panel correspond to a mean altitude of the line-of-sight of 105 km and those in the lower panel to a mean altitude of 145 km. The error bars correspond to the 1-SD Noise Equivalent Spectral Radiance. The CIA from the $S_0(0)$ and $S_0(1)$ H$_2$ lines is not visible in these limb spectra that probe the stratosphere.

*2.2 Radiative transfer model*

Synthetic spectra were generated using a line-by-line radiative transfer code. The opacity sources we included are the N$_2$-N$_2$, N$_2$-CH$_4$, N$_2$-H$_2$ and CH$_4$-CH$_4$ collision-induced absorption (CIA), absorption by aerosol particles and molecular lines from CH$_4$, C$_2$H$_2$, C$_2$H$_4$, C$_2$H$_6$, $CH_3C_2H$, $C_4H_2$, $C_3H_8$, HCN, CO and $CO_2$. These molecules contribute through their bands at 70-160 $cm^{-1}$ (rotational lines of $CH_4$), 729 $cm^{-1}$ ($C_2H_2$), 949 $cm^{-1}$ ($C_2H_4$), 822 $cm^{-1}$ ($C_2H_6$), 327 and 633 $cm^{-1}$ ($CH_3C_2H$), 220 and 628 $cm^{-1}$ ($C_4H_2$), 748 $cm^{-1}$ ($C_3H_8$), < 60 $cm^{-1}$ (rotational lines of CO and HCN), 713 $cm^{-1}$ (HCN) and 667 $cm^{-1}$ ($CO_2$).

To compare with FP1 limb spectra, synthetic spectra calculated on a fine (10-km) grid of altitudes were convolved averaging the FOV corresponding to the altitudes and distances of the individual spectra in the 105-km or 145-km FP1 selection. The FP1 FOV was approximated by a Gaussian function up to 3.2 mrad from center with a half width at half maximum of 1.2 mrad. This function reproduces quite well the FOV sensitivity function derived by Anderson and Samuelson (2011, Fig. 2). For FP3 limb spectra, which have a much smaller FOV, we just weighted the three synthetic spectra included in the altitude range chosen for each selection (Col. 5 of Table 1) according to the distribution in altitude of the individual CIRS spectra.

Spectroscopic line parameters are the same as in Bézard et al. (2018). The CIA coefficients were calculated from the codes available at https://www.astro.ku.dk/~aborysow/programs/ and based on modelling by Borysow and Frommhold (1986a, 1986b, 1987) and Borysow and Tang (1993). However, previous analyses of CIRS FP1 spectra led to the conclusion that the $N_2$-$CH_4$ CIA coefficients are probably underestimated by some 50% at the low temperatures of Titan's upper troposphere (70-85 K) (Tomasko et al. 2008a, de Kok et al. 2010). Anderson and Samuelson (2011) also concluded that the coefficients of Borysow and Tang are in error but used a wavenumber-dependent correction factor $\beta_\nu$ rather than the constant factor of 0.5 suggested by Tomasko et al. and de Kok et al. In this study, as did Anderson and Samuelson, we multiplied the $N_2$-$CH_4$ CIA coefficients by:

$$CF_{N_2-CH_4} = 1 + \frac{\beta}{(T-T_{sat})^2+1}, \qquad (2)$$

where $T$ and $T_{sat}$ are air temperature and CH$_4$ saturation temperature but we assumed that $\beta$ does not vary with wavenumber. We will come back to this point in Section 2.4.

In the case of N$_2$-N$_2$ absorption, we have also considered the more recent quantum mechanical calculations of Karman et al. (2015), who used ab initio dipole moment and potential energy surfaces as well as the full anisotropic interaction potential. In contrast, Borysow and Frommhold (1986a) used effective isotropic potentials combined with an adjusted short range induced dipole moment. While both calculations agree within 10% between 50 and ~110 cm$^{-1}$, the far wing absorption calculated by Karman et al. is significantly larger than that of Borysow and Frommhold, especially at low temperatures. The disagreement reaches a factor of 2 at 180 cm$^{-1}$ and 78 K. We have thus multiplied the Borysow and Frommhold CIA coefficients beyond 110 cm$^{-1}$ by a temperature- and wavenumber-dependent factor coefficient:

$$CF_{N_2-N_2} = 2^{\frac{\sigma-110}{2.5 \times T-100}}, \qquad (3)$$

with $\sigma$ in cm$^{-1}$ and $T$ in K. Doing so, our coefficients reproduce Karman et al's calculations (given in their Table III) within 10% in the ranges $\sigma$ = 45-180 cm$^{-1}$ and $T$ = 70-100 K. We compare in Section 2.4.2 calculations of Titan's spectrum performed with Borysow and Frommhold's raw coefficients and with this correction factor.

To calculate the main haze opacity, we utilized the spectral dependence of the extinction cross section derived from Cassini/CIRS limb measurements by Vinatier et al. (2012) from 630 to 1500 cm$^{-1}$ and by Anderson and Samuelson (2011) below 560 cm$^{-1}$. As discussed in Section 2.4, the coefficients of Anderson and Samuelson produce spurious features in Titan's 105- and 145-km limb spectra and we had to smooth them. We added the opacity of the nitrile haze characterized by Anderson and Samuelson (2011), which peaks at 160 cm$^{-1}$ and is significant from 90 to 290 cm$^{-1}$ as shown in their Fig. 15.

*2.3 Atmospheric model*

The methane mole fraction profile is the one proposed by Rey et al. (2018), equal to the Huygens/GCMS profile below 39 km (Niemann et al. 2010), decreasing with altitude at a rate of $7.5 \times 10^{-5}$ per km up to 110 km to reach a value of 0.010 above 110 km. This profile allows us to reproduce the absorption of direct solar flux in different methane bands as measured by the Huygens/DISR. A stratospheric value of ~0.010 at low latitudes, significantly lower than the Huygens/GCMS mole fraction (0.0148), also agrees with the analysis of methane rotational lines from Cassini/CIRS data by Lellouch et al. (2014). The hydrogen mole fraction is held constant with altitude. The H$_2$ para fraction $f_{para}$ is either at local equilibrium or constant with altitude.

We used the vertical profiles of $C_2H_6$, $C_2H_2$, $C_2H_4$, $CH_3C_2H$, $C_4H_2$, $C_3H_8$, and HCN derived by Vinatier et al. (2010a) from limb measurements near 21°S in March 2007[1]. The $CO_2$ and CO mole fractions were held constant at $1.6 \times 10^{-8}$ and $4.7 \times 10^{-5}$ respectively (de Kok et al. 2007). We allowed for small multiplicative factors, between 0.75 and 1.2, that we applied to the vertical profiles of $C_2H_6$, $C_2H_2$, $C_2H_4$, $CH_3C_2H$ and HCN to best reproduce the emission bands in our selections.

For the main haze profile, we used the extinction profile of Doose et al. (2016) at 1 μm up to 152 km and assumed a constant scale height of 45 km above. We also modified the *b* and *h* of their model parameters (Table 2 of Doose et al.) from 6 to 5.8 and from 1.35 to 1.40 respectively. This small modification allowed us to bring the 105- and 145-km FP3 and FP1 continuum emission in much better agreement. The only free parameter left is a scaling factor between the 1-μm extinction and that at 1090 cm$^{-1}$, that we chose as a reference wavenumber. Its value is ~ 0.011 (Bézard et al. 2018, Fig. 2).

---

[1] Temperature and abundance profiles available from the Virtual European Solar and Planetary Access (VESPA) http://vespa.obspm.fr/planetary/data/

The temperature profile is based on the Huygens/HASI profile (Fulchignoni et al. 2005) complemented by information from Cassini/CIRS spectra above ~100 km. It was retrieved from inversion of radiances in the methane band between 1210 and 1319 cm$^{-1}$ as described in Vinatier et al. (2007) and using the CH$_4$ vertical profile described above. A smoothed version of the Huygens/HASI temperature profile served as the a priori profile in the inversion process. We selected CIRS FP4 limb and nadir spectra recorded at 0.5-cm$^{-1}$ resolution during Flyby TB (13 December 2004), one month before the Huygens descent, using the same criteria as Vinatier et al. (2007) and Lellouch et al. (2014). The nadir selection corresponds to a mean latitude of 8.8°S and a mean emission angle of 35°. The eight limb selections at a latitude of 13°S span the altitude range 230-480 km. Figure 1 shows the retrieved profile along with the raw HASI profile. The comparison shows a very good agreement in the altitude range 100-147 km where information from both the CIRS nadir selection and the temperature sensor of HASI is available, as previously noted by Lellouch et al. (2014). On the other hand, retrieving temperatures from CIRS measurements with the Huygens/GCMS CH$_4$ profile leads to temperature 3-5 K colder than the HASI direct measurements between 80 and 147 km (Lellouch et al. 2014).

*2.4 Results*

*2.4.1 Aerosol opacity*

We began our analysis with the 145-km limb spectrum, which is not affected by collision-induced absorption, in order to constrain the wavenumber dependence of the main haze opacity. To best reproduce the emission bands of $C_2H_2$, $C_2H_6$, $CH_3C_2H$, $C_2H_4$ and HCN, we multiplied our nominal abundance profiles by factors of 0.80, 0.90, 0.80, 0.80 and 1.20 respectively. Profiles of other molecules were left unchanged. We used the haze spectral extinction from Vinatier et al. (2012) above 630 $cm^{-1}$ normalized to a normal haze optical depth of 0.032 at 1090 $cm^{-1}$ and reduced their 610-$cm^{-1}$-value by 7% to reproduce the FP3 pseudo-continuum. Below 600 $cm^{-1}$, we adjusted the haze extinction to reproduce the FP1 continuum. To do so, we had to smooth spectrally the Vinatier et al. coefficients, but essentially within their error bars. The Vinatier et al. coefficients, derived from the analysis of FP1 limb spectra by Anderson and Samuelson (2011), are affected by instrumental noise, which causes spurious features not seen in our large averages of FP1 spectra. In addition, our extinction coefficients below 610 $cm^{-1}$ are overall 5-10% larger than those of Vinatier et al. with respect to those beyond 630 $cm^{-1}$ (Fig. 4). Our best fit model is shown in Fig. 5.

We then turned to the 105-km limb spectrum, which we modeled with the same haze properties as for the 145-km spectrum. The factors by which we multiplied the $C_2H_2$ and $C_2H_6$ reference abundance profiles are 0.75 and 0.85 respectively (vs. 0.80 and 0.90 for the 145-km spectrum). For the other compounds, the factors are the same as for the 145-km spectrum. The 105-km average spectrum is sensitive to the CIA below ~320 $cm^{-1}$, due to the altitude range of our selection (90-120 km) and the large extent of the FP1 FOV, which all together imply a significant contribution from the lower stratosphere. The synthetic spectrum is shown in Fig. 5 (dashed line). In this calculation, the $N_2$-$N_2$ and $N_2$-$CH_4$ Borysow and Frommhold (1986a) and Borysow and Tang (1993) CIA coefficients have been corrected as described in Section 2.2 (with $\beta = 0.5$ for $N_2$-$CH_4$ in Eq. 2). This spectrum underestimates the observed emission in the 100-300 cm$^{-1}$ range as originally pointed out by Anderson and Samuelson (2011). We then added the emission from the nitrile ice cloud modeled by these authors and found that a normal optical depth of 0.0057 at 160 cm$^{-1}$ best reproduces our observations (solid line in Fig. 5). This value is similar to that derived by Anderson and Samuelson at 15°S. Note that if we use the Borysow and Frommhold (1986a) $N_2$-$N_2$ CIA coefficients as did these authors, we need an optical depth of 0.0065 (at 160 cm$^{-1}$), intermediate between their values at 15°S (0.0054) and 15°N (0.0107).

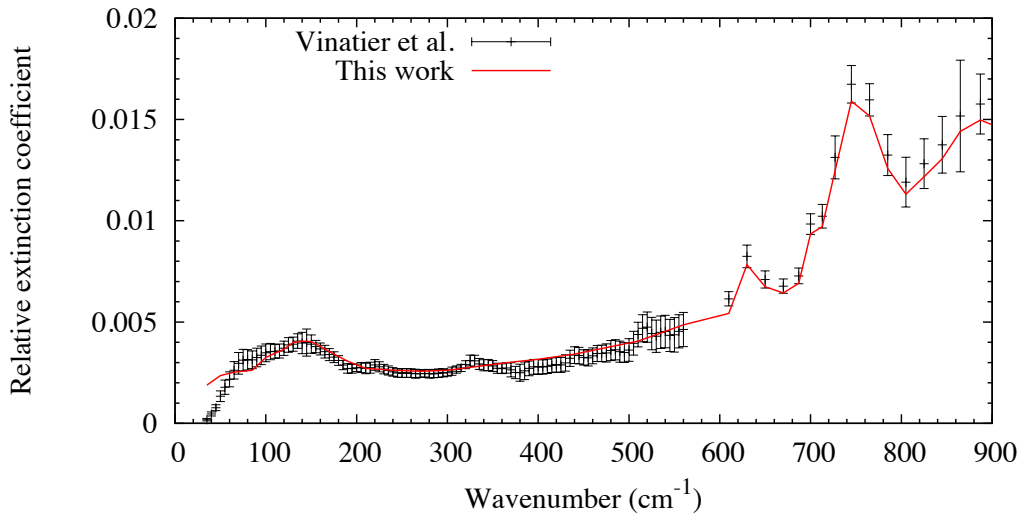

*2.4.2 $N_2$-$CH_4$ and $N_2$-$N_2$ collision-induced absorption*

**Fig. 4.** Spectral dependence of the main haze opacity. Solid line represents the haze spectral opacity in our model relative to that of Doose et al. (2016) at 1-µm wavelength. Dots with 1-SD error bars represent the spectral dependence of the haze extinction inferred by Vinatier et al. (2012) and normalized (arbitrarily) to our model at 540 cm$^{-1}$.

Having fixed the aerosol (main haze and nitrile ice cloud) spectral opacity, we analyzed the spectral selections at low (13°) and high (59°) emission angles shown in Fig. 2. The opacity below 600 cm$^{-1}$ is dominated by CIA from $N_2$-$N_2$, $N_2$-$CH_4$ and $N_2$-$H_2$ mostly of tropospheric origin, because of its density-squared dependence. In the stratosphere, besides aerosol opacity, $CH_3C_2H$ and $C_4H_2$ add some emission around 327 and 220 cm$^{-1}$. For these molecules, we used the abundance profiles that reproduce the emission bands in the limb spectra (Fig. 5) and also the 630-cm$^{-1}$ emission feature in nadir spectra (Figs. 8-9).

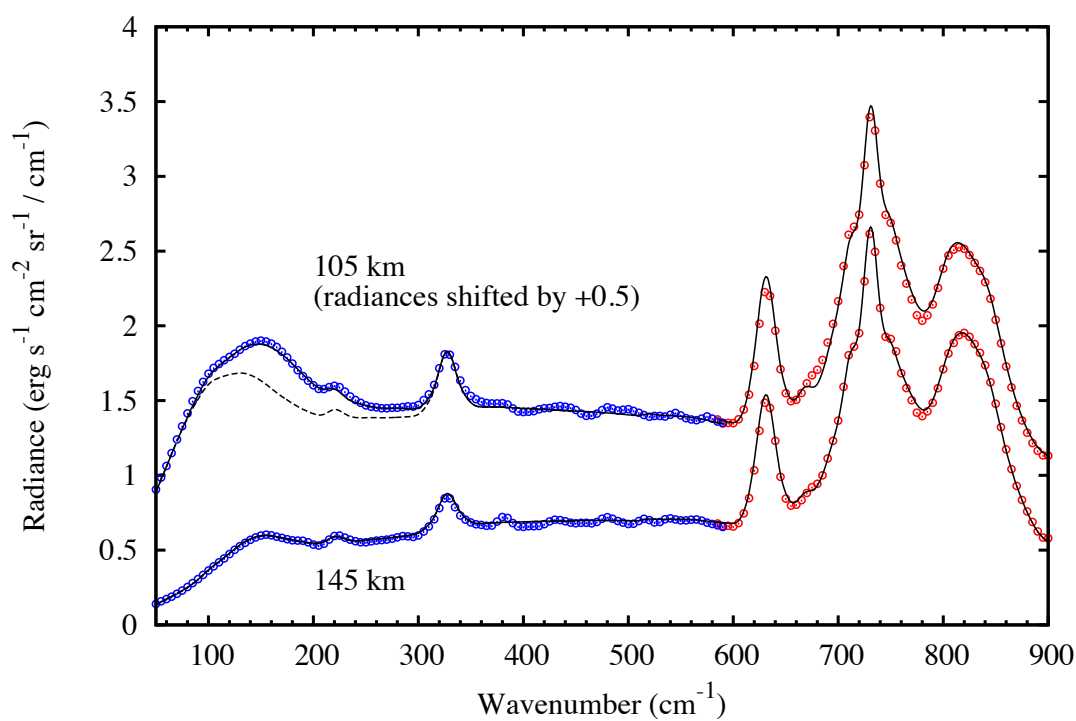

**Fig. 5.** FP1 (blue circles) and FP3 (red circles) CIRS average limb spectra at 105 and 145 km are compared with synthetic spectra using the same haze model with the spectral dependence shown in Fig. 4 (dashed lines). Solid lines represent our best fit model in which opacity from a nitrile ice cloud, as parametrized by Anderson and Samuelson (2011), is added.

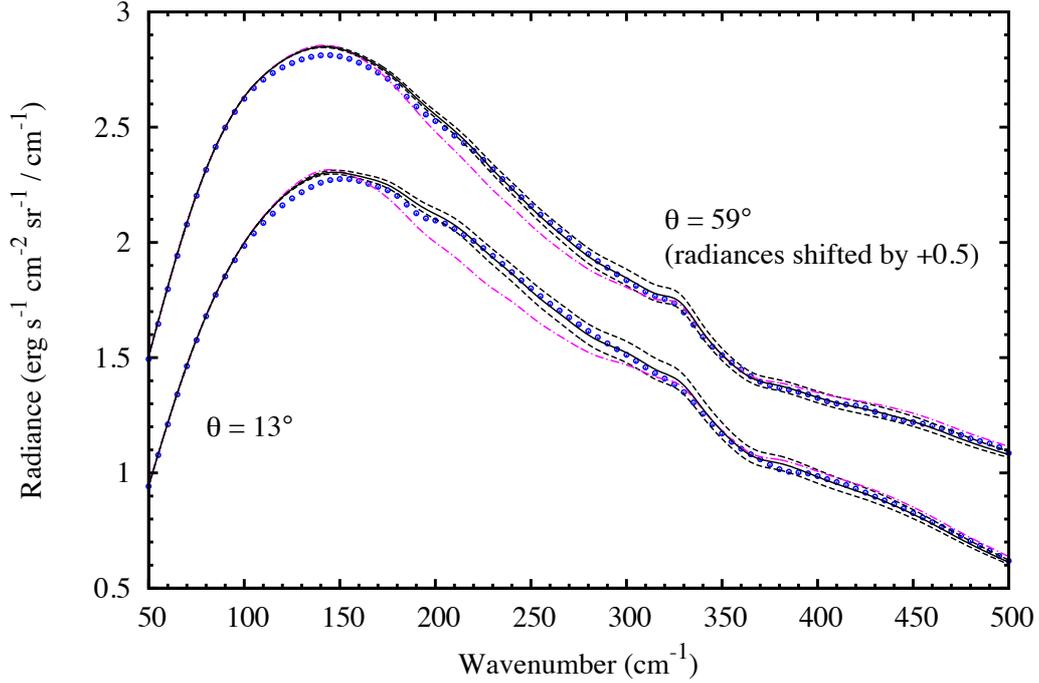

**Fig. 6.** FP1 low (13°) and high (59°) emission angle spectra (blue circles) are compared with synthetic spectra using $ß = 0.5$ (solid line, nominal case), 0.3 and 0.7 (dashed lines). Increasing $ß$ lowers the calculated emission for both emission angles. Also shown is a calculation using $ß$ from Anderson and Samuelson (2011) (magenta dash-dotted line). The high-emission angle spectra are shifted by 0.5 erg s$^{-1}$ cm$^{-2}$ sr$^{-1}$ / cm$^{-1}$ for clarity.

We first considered the $N_2$-$CH_4$ and $N_2$-$N_2$ CIA. As mentioned above, the CIA calculated from Borysow and Tang's (1993) coefficients is not strong enough to reproduce Titan's emission between ~180 and 460 cm$^{-1}$. We show in Fig. 6 the effect of varying the factor $ß$ by which we increase the $N_2$-$CH_4$ coefficients at altitudes where methane is saturated (Eq. 2). The best fitting value, which minimizes the residuals between 180 and 325 cm$^{-1}$, is 0.52 for both CIRS spectra. We assign an uncertainty of ± 0.05 to this parameter. Increasing $ß$ lowers the calculated emission for both emission angles. Beyond 325 cm$^{-1}$, $N_2$-$H_2$ CIA becomes important. Though it is quite possible that $ß$ varies with wavenumber as suggested by Anderson and Samuelson (2011), we find that a constant value reproduces satisfactorily the CIRS spectra (to better than 1%). On the other hand, the large spectral variation of $ß$ suggested by Anderson and Samuelson (2011, Fig. 13), essentially from low-altitude limb spectra, leads to synthetic radiances that are at odds with the observations: much too low in the 180-300 cm$^{-1}$ interval and a bit too high around 400-450 cm$^{-1}$ (Fig. 6). We also made a test calculation in which we replaced our CH$_4$ profile (Rey et al. 2018) with that of Niemann et al. (2010) and found a maximum radiance variation of 0.1% around 100-125 cm$^{-1}$. This variation is negligible because the two mixing ratio profiles differ only in the stratosphere while the CIA, varying as the squared density, mostly takes place in the troposphere.

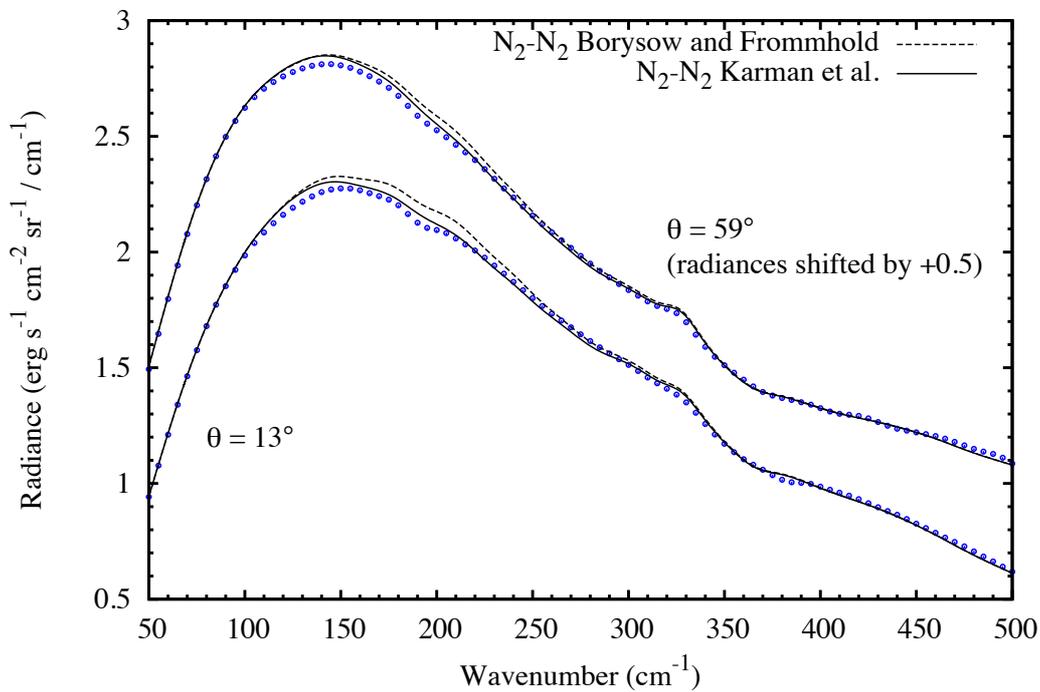

**Fig. 7.** FP1 low (13°) and high (59°) emission angle spectra (blue circles) are compared with synthetic spectra using N$_2$-N$_2$ CIA coefficients from Borysow and Frommhold (1986a) (dashed lines) and with the corrective factor of Eq. 3 to reproduce Karman et al.'s (2015) ab initio calculations (solid lines). $\beta = 0.52$ is used to correct N$_2$-CH$_4$ CIA (Eq. 2). The high-emission spectra are shifted by 0.5 erg s$^{-1}$ cm$^{-2}$ sr$^{-1}$ / cm$^{-1}$ for clarity.

Our modelling of the N$_2$-N$_2$ CIA is based on the ab initio calculations of Karman et al. (2015), that we used to define a multiplicative corrective factor to the Borysow and Frommhold (1986a) CIA coefficients longward of 110 cm$^{-1}$ (Eq. 3). Spectra calculated with the raw Borysow and Frommhold coefficients clearly overestimate the observed emission between approximately 110 and 250 cm$^{-1}$ (Fig. 7). The disagreement is a bit larger at low emission angle. This points to a missing opacity in the troposphere since reducing the stratospheric haze optical depth would decrease the radiances more efficiently at high than at low emission angle. Increasing the N$_2$-N$_2$ CIA with our corrective factor (and accordingly reducing the nitrile ice optical depth from 0.0064 to 0.0057 to still fit the 105-km spectrum) significantly reduces the discrepancy beyond 140 cm$^{-1}$. Our calculations still slightly overestimate the observed emission from 110 to 200 cm$^{-1}$, by at most 2% around 120-150 cm$^{-1}$.

*2.4.3 H$_2$ abundance and para fraction*

We finally investigated the H$_2$ mole fraction and ortho-to-para ratio from fitting of the 300-650 cm$^{-1}$ interval, which is affected by the N$_2$-H$_2$ CIA through the $S_0(0)$ and $S_0(1)$ lines at 355 and 590 cm$^{-1}$. The C$_2$H$_2$ and C$_2$H$_6$ abundance profiles are our nominal ones multiplied by 0.90 and 0.95 to reproduce the emission bands at 729 and 822 cm$^{-1}$. The profiles of the other molecular species and of the aerosol opacity are the same as those used to model the two limb spectra. As discussed just above, the N$_2$-CH$_4$ and N$_2$-N$_2$ CIA coefficients were modified from the calculations of Borysow and Tang (1993) and of Borysow and Frommhold (1986a) following Eq. 2 (with $\beta = 0.52$) and Eq. 3 respectively.

To derive the best-fitting parameters, we minimized the chi-squared ($\chi^2$) between the observations and the calculated spectra for the spectral selections at low and high emission angles. For the error used to calculate the $\chi^2$, we combined the standard deviation of the selection divided by $\sqrt{N}$, where $N$ is the number of spectra, and a model error that we set as $0.008 \times I_\nu + 0.003$ erg s$^{-1}$ cm$^{-2}$ sr$^{-1}$ / cm$^{-1}$ where $I_\nu$ is the observed spectral radiance. This model error thus increases from ~1% of the radiance at 300 cm$^{-1}$ to ~2% at 600 cm$^{-1}$. We consider that this error is realistic considering that we were able to fit the 180-325 cm$^{-1}$ interval, outside of the H$_2$ lines, with a precision of about 1%, and also, because it yields minimum values of the reduced $\chi^2$ of about 1 in the regions of both the $S_0(0)$ and $S_0(1)$ lines (300-450 cm$^{-1}$ and 470-650 cm$^{-1}$ respectively). The best fitting H$_2$ mole fraction and para fraction correspond to a minimum of the reduced $\chi^2$ ($\chi_r^2 = \chi_{r,min}^2$) and 1-SD error bars on these parameters correspond to $\chi_r^2 = \chi_{r,min}^2 \times \frac{n_p+1}{n_p}$, where $n_p$ is the number of independent spectral points in the two selections. We take $n_p = 2\frac{\Delta\sigma}{\delta\sigma} - n_m$, where $\Delta\sigma$ is the width of the spectral interval of the fit (350-650 cm$^{-1}$), $\delta\sigma$ is the spectral resolution (15.9 cm$^{-1}$) and $n_m$ the number of independent points in the high emission angle selection where the model error is larger than the noise error (350-450 cm$^{-1}$), considering that any model error is mostly correlated between low and high emission spectra. This yields $n_p = 31$.

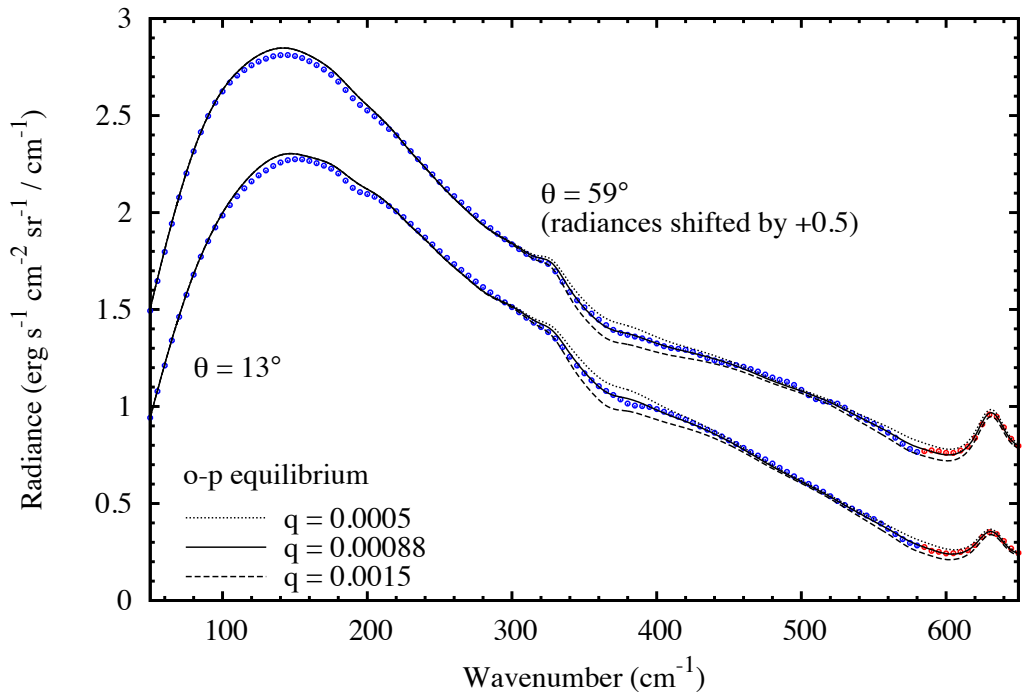

**Fig. 8.** Low (13°) and high (59°) emission angle FP1 (blue circles) and FP3 (red circles) spectra are compared with synthetic spectra assuming o-p H$_2$ equilibrium and uniform H$_2$ mole fractions of 0.5, 0.88 and 1.5 × 10$^{-3}$. The high-emission spectra are shifted by 0.5 erg s$^{-1}$ cm$^{-2}$ sr$^{-1}$ / cm$^{-1}$ for clarity.

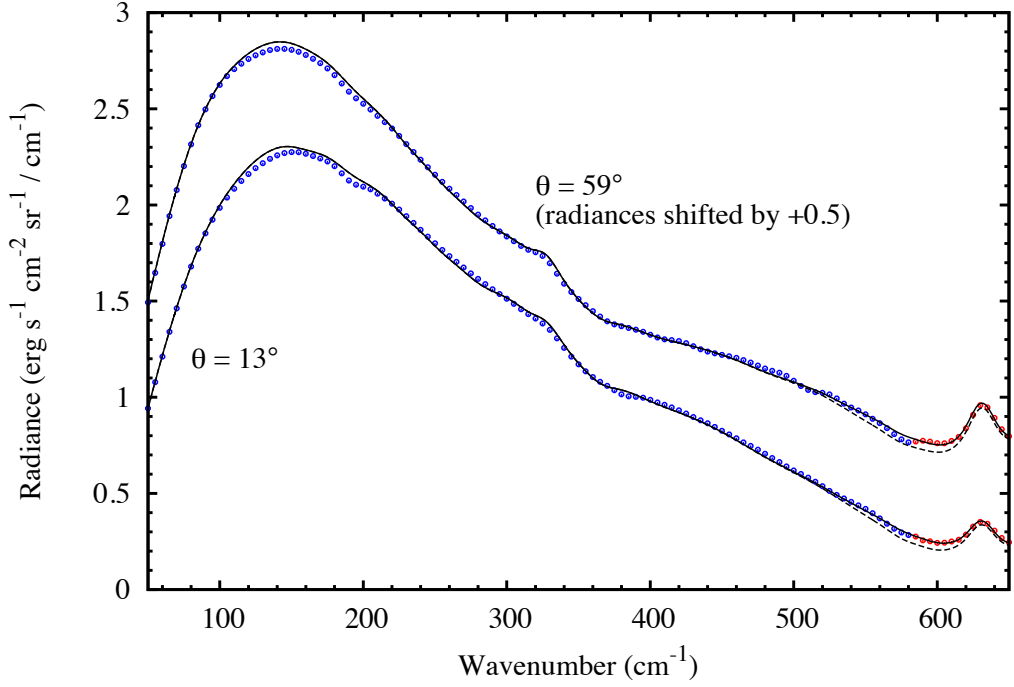

**Fig. 9.** Low (13°) and high (59°) emission angle FP1 (blue circles) and FP3 (red circles) spectra are compared with synthetic spectra assuming a uniform H$_2$ para fraction of 0.51 with a H$_2$ mole fraction of $0.89 \times 10^{-3}$ (solid line), or a uniform H$_2$ para fraction of 0.35 with a H$_2$ mole fraction of $1.3 \times 10^{-3}$ (dashed line). The high-emission spectra are shifted by 0.5 erg s$^{-1}$ cm$^{-2}$ sr$^{-1}$ / cm$^{-1}$ for clarity.

We first assumed that the o-p ratio is at equilibrium at all levels. In this case, we inferred a H$_2$ mole fraction of $(0.875 \pm 0.03) \times 10^{-2}$ with an associated reduced chi-squared ($\chi_r^2$) of 0.97. The corresponding spectrum reproduces quite well the observations as shown in Fig. 8 (solid line). The fact that the low and high emission angle spectra are reproduced equally well indicates that the tropospheric and stratospheric contributions are well accounted for, especially the stratospheric haze emission. Also shown are calculations with H$_2$ mole fractions of 0.5 and $1.5 \times 10^{-3}$ to illustrate the sensitivity of the spectrum calculations to the H$_2$ abundance.

In a second step, we investigated a case in which the o-p ratio is not at equilibrium at all levels but instead constant with altitude. Doing so, the best-fitting $H_2$ para fraction ($f_{para}$) is $0.51 \pm 0.02$ and the $H_2$ mole fraction $(0.89 \pm 0.03) \times 10^{-3}$, with a minimum value of $\sqrt{\chi_r^2}$ equal to 1.00. This value is only marginally larger (at the 1-SD confidence level) than in the equilibrium case and we can thus consider that a uniform $f_{para}$ value of 0.51 reproduces equally well the observations. This is illustrated in Fig. 9, where we also show a spectrum with $f_{para} = 0.35$ and a $H_2$ mole fraction of 0.13%, so that the para-$H_2$ abundance is unchanged. The $S_0(0)$ absorption near 355 cm$^{-1}$ is thus equally well reproduced but, for $f_{para} = 0.35$, the radiance around the $S_0(1)$ line centered at 590 cm$^{-1}$ is much lower than observed.

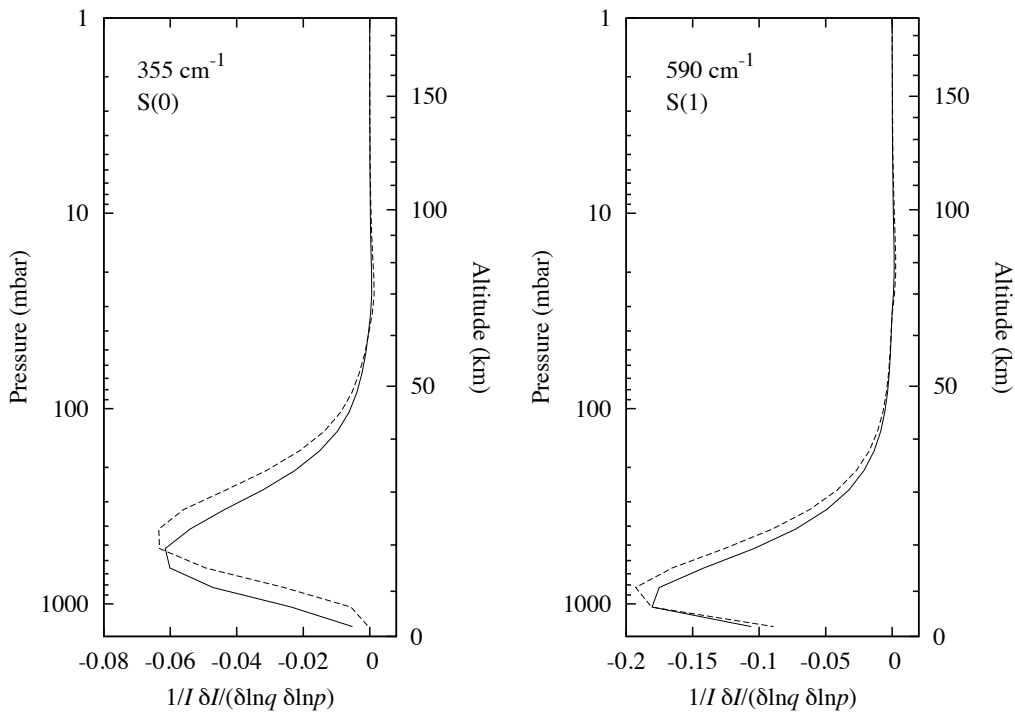

**Fig. 10.** Sensitivity of the spectral radiance at 355 (left panel) and 590 (right panel) cm$^{-1}$ at low (13°, solid lines) and high (59°, dashed lines) emission angles to a local change in the $H_2$ mole fraction from the nominal value of $0.89 \times 10^{-3}$, assuming o-p equilibrium. The quantity shown is the relative variation of the spectral radiance ($\delta I/I$) for an infinitesimal relative change of the $H_2$ mole fraction $\delta \ln q$ in a layer of height $\delta \ln p$, divided by $\delta \ln q \; \delta \ln p$.

Figure 10 shows the sensitivity of the radiance at 355 and 590 cm$^{-1}$ to the H$_2$ mole fraction as a function of altitude for the low and high emission angles. In the core of the $S_0$(0) line, the sensitivity peaks at 18 km (resp. 21 km) with a width at half maximum of 8-31 km (resp. 12-34 km) at low (resp. high) emission angle. The $S_0$(1) line probes the H$_2$ abundance a bit deeper, 8 km (resp. 9 km) with a half-maximum sensitivity range of 1-20 km (resp. 3-22 km) at low (resp. high) emission angle.

In conclusion, we do not see evidence for departure from H$_2$ o-p equilibrium in the region probed by the H$_2$ lines, mostly below 34 km. However, we cannot discard the possibility of a uniform H$_2$ para fraction of 0.49-0.53 in this region as such a model yields an almost as good reproduction of the observed spectra. In any of these cases, the best-fitting H$_2$ mole fraction lies between 0.84 and 0.92 × 10$^{-3}$ at the 1-SD confidence level. This result pertains to the equatorial region in the 2005-2007 period.

## 3. H$_2$ ortho-para conversion in Titan's atmosphere

*3.1 Conversion mechanisms*

Interconversion of para and ortho hydrogen in the Jovian planets was first investigated by Massie and Hunten (1982) and later, in more detail, by Conrath and Gierasch (1984) and Fouchet et al. (2003). Processes that can result in ortho-para (o-p) equilibration are (1) hydrogen exchange in the gas phase, (2) magnetic interaction in the gas phase, and (3) interaction with surfaces in a physisorption state. These processes are also relevant to Titan and we review them below.

*3.1.1. Hydrogen exchange in the gas phase*

Collisions between molecular and atomic hydrogen can lead to exchange of H and equilibration of ortho and para states: p-$H_2$ + H → o-$H_2$ + H (R1a) or conversely o-$H_2$ + H → p-$H_2$ + H (R1b). The characteristic time for equilibration is then given by:

$$\tau_{exch} = 1/[n_H(k_{p\to o} + k_{o\to p})], \quad (4)$$

where $n_H$ is the H atom number density, $k_{p\to o}$ the reaction rate of (R1a) and $k_{o\to p}$ the reaction rate of (R1b). It can be written as:

$$\tau_{exch} = (1 - f_{para}^{eq})/(n_H\, k_{p\to o}), \quad (5)$$

where $f_{para}^{eq}$ is the equilibrium $H_2$ para fraction. Schatz and Kuppermann (1976) have presented an accurate quantum mechanical calculation of the reaction rate $k_{p\to o}$ as a function of temperature. We used $k_{p\to o}$ = 4.0 × $10^{-12}$ exp(-2600/$T$) + 1.1 × $10^{-15}$ exp(-1240/$T$) $cm^3$ $molecule^{-1}$ $s^{-1}$ (with $T$ expressed in K), which reproduces their calculations to better than 10% over the range 100-300 K.

We used the H number density profile from the model of Vuitton et al. (2019) that includes removal of H atoms by chemisorption at the surface of aerosols, mostly in the region 200-450 km. We also tested the H number density profile from Lavvas et al. (2008) that includes heterogeneous reactions of H at the surface of aerosols. This profile exhibits a ~20 times larger H density at 300-400 km than that of Vuitton et al. (2019). The so-calculated profile of $\tau_{exch}$ is shown in Fig. 11.

*3.1.2. Magnetic interaction in the gas phase*

Collisions with paramagnetic molecules in the gas phase can induce o-p transitions in $H_2$. However, all molecules in Titan's atmosphere are diamagnetic except for radicals, such as H or CH3, which are not abundant enough to play a significant role in this regard. Diamagnetic molecules do not have unpaired electrons that would produce a permanent magnetic moment from their orbital angular momenta. However, a much weaker magnetic moment arises from the spin of the molecule nuclei. For example, ortho-H2 has a total nuclear spin $I = 1$ and a magnetic moment $\mu_{oH_2} = \mu_n\, g_p\, I$, where $\mu_n$ is the nuclear magneton and $g_p$ the g-factor of a proton (5.586). Given that the o-p conversion during a collision scales as the square of the magnetic moment of the perturber, Conrath et al. (1984) extrapolated the o-p H2 conversion probability measured for O2 collisions at 125 K to that for collisions with ortho-H2. The O2 molecule is paramagnetic with a magnetic moment $\mu = \sqrt{8}\,\mu_b$, where $\mu_b$ is the Bohr magneton ($\mu_b = 1836\,\mu_n$), implying that the conversion probability with ortho-H2 is almost $10^6$ times weaker than with O2. In 1997, Milenko et al. published measurements of the ortho-to-para hydrogen conversion rate at 40-120 K in the gas phase. Their data can be represented in the form:

$$k_{o \to p} = 2.5 \times 10^{-28} \left(\frac{T}{125}\right)^{0.56} \text{cm}^3 \text{ molecule}^{-1}\text{ s}^{-1} \ (T \text{ expressed in K}), \qquad (6)$$

which is only 4 times larger at 125 K than the estimation of Conrath et al. and indicates that their extrapolation from the oxygen case was relatively valid. Milenko et al. (1997) interpreted their data with a theory that takes into account the dependence of the approach distance between colliding molecules on velocity. Using the Lennard-Jones potential for the intermolecular interaction, their model yields the correct $\sim T^{7/12}$ temperature dependence of $k_{o \to p}$, contrary to Wigner's theory that was used by Conrath et al. and predicts a $T^{-1/2}$ dependence. According to Milenko et al.'s model, $k_{o \to p}$ varies as $\varepsilon^{-7/12}\,\sigma^{-5}$ where $\varepsilon$ and $\sigma$ are the parameters of the Lennard-Jones potential (respectively the depth of the potential well and the distance at which the inter-particle potential is zero), assuming that the collision frequency varies as $\sigma^2$.

However, in contrast with giant planets, hydrogen is not abundant enough in Titan's atmosphere to contribute significantly to this conversion mechanism, and we consider instead the more abundant $N_2$ and $CH_4$ molecules. The two $I = 1$ spins of the N atoms can combine to yield a resulting spin in $N_2$ equal to $I = 0$ with a weight of 1, $I = 1$ with a weight of 3, or $I = 2$ with a weight of 5. The average squared magnetic moment of a $N_2$ molecule is thus

$$\mu_{N_2}^2 = \frac{3\,(1)^2 + 5\,(2)^2}{9} g_N^2 \mu_n^2 = 0.417\, \mu_n^2, \tag{7}$$

where $g_N = 0.4038$ is the nuclear $g$-factor of the N atom (i.e. the nuclear magnetic moment of the N atom is $g_N I$, in units of the nuclear magneton $\mu_n$). In the methane molecule, the magnetic moment of the C atom is zero while the four $I = 1/2$ spins of the H atoms couple to give a total spin of $I = 0$ with a weight of 2, $I = 1$ with a weight of 9, or $I = 2$ with a weight of 5. The average squared magnetic moment of a $CH_4$ molecule is thus equal to:

$$\mu_{CH_4}^2 = \frac{9\,(1)^2 + 5\,(2)^2}{16} g_p^2 \mu_n^2 = 56.5\, \mu_n^2, \tag{8}$$

where $g_p = 5.586$ is the nuclear $g$-factor of the proton. These values have to be compared to that for an ortho-$H_2$ molecule, $\mu_{oH_2}^2 = g_p^2 \mu_n^2 = 31.2\, \mu_n^2$, to scale the reaction rate of Eq. 6 for collisions with $N_2$ or $CH_4$ molecules. It is also necessary to account for the differences in the Lennard-Jones parameters between $H_2$-$H_2$, $N_2$-$H_2$ and $CH_4$-$H_2$. For the last two mixtures, we calculate $\sigma$ as the arithmetic mean of those for single species and $\varepsilon$ as the harmonic mean of those for single species. Doing so, the $\varepsilon^{-7/12} \sigma^{-5}$ scaling factor mentioned above is respectively 0.44 and 0.39 time smaller for $N_2$ and $CH_4$ colliders than for $H_2$. We finally obtain the following ortho-to-para $H_2$ conversion rates from scaling of Eq. 6:

in $N_2$: $k_{o \to p}^{N_2} = 1.5 \times 10^{-30} \left(\frac{T}{125}\right)^{0.56}$ cm³ ($N_2$ molecule)⁻¹ s⁻¹ ($T$ expressed in K), $\qquad$ (9)

in $CH_4$: $k_{o \to p}^{CH_4} = 1.7 \times 10^{-28} \left(\frac{T}{125}\right)^{0.56}$ cm³ ($CH_4$ molecule)⁻¹ s⁻¹ ($T$ expressed in K). $\qquad$ (10)

The conversion time in Titan's atmosphere is then given by:

$$\tau_{gas} = 1/\left[n_{N_2}\left(k_{o \to p}^{N_2} + k_{p \to o}^{N_2}\right) + n_{CH_4}\left(k_{o \to p}^{CH_4} + k_{p \to o}^{CH_4}\right)\right], \tag{11}$$

where $n_{N_2}$ and $n_{CH_4}$ are respectively the N$_2$ and CH$_4$ number densities. This yields:

$$\tau_{gas} = f_{para}^{eq} / \left[ n \left( q_{N_2} k_{o \to p}^{N_2} + q_{CH_4} k_{o \to p}^{CH_4} \right) \right], \qquad (12)$$

where $f_{para}^{eq}$ is the equilibrium H$_2$ para fraction, $n$ the atmospheric number density (in molecule cm$^{-3}$), $q_{N_2}$ the N$_2$ mole fraction, $q_{CH_4}$ the CH$_4$ mole fraction, $k_{o \to p}^{N_2}$ the reaction rate in Eq. 9 and $k_{o \to p}^{CH_4}$ the reaction rate in Eq. 10. The vertical profile of $\tau_{gas}$ is given in Fig. 11.

### 3.1.3. Magnetic interaction on aerosol surface

When molecular hydrogen is adsorbed on an aerosol particle, interconversion between the ortho and para states can occur through magnetic interaction with the surface material. Note that, although this mechanism is most effective for magnetic materials, it can also work for diamagnetic materials (Fukutani and Sugimoto 2013). The first step in this process is molecular adsorption, which occurs through the Van der Waals interaction. Key parameters in this regard are the collision rate of H$_2$ molecules with aerosols ($C$) and the sticking probability ($S$). In a physisorption state, the o-p conversion time ($\tau_c$) and the residence ($t_r$) time on the surface are of course important, thermal desorption being in competition with the conversion process at the surface. The residence time is limited by the desorption rate ($t_r = 1/k_{des}$). Another parameter is the surface density of available sites ($n_s$) which limits the adsorption rate (the molecule is not adsorbed if it hits an occupied site). The o-p conversion time in Titan's atmosphere is then given by:

$$\tau_{aer} = \frac{\tau_c \, k_{des}}{S \, C} f_{sites}, \qquad (13)$$

where $f_{sites}$ is the ratio of total site density to non-occupied site density. In steady state, the adsorption and desorption rates on a particle are equal, which gives $f_{sites} = 1 + \frac{S \, n \, v}{4 \, k_{des} \, n_s}$. Here $n$ is the atmospheric number density, $v = \sqrt{\frac{8kT}{\pi m}}$ is the mean collision speed, $m$ is the mean atmospheric mass, assuming that all atmospheric molecules can be adsorbed with the same efficiency *S*. Some of these parameters are uncertain and the following discussion is based on the paper reviewing o-p conversion of $H_2$ on solid surfaces by Fukutani and Sugimoto (2013) and on the investigation of this mechanism on interstellar grains by Bron et al. (2016).

Assuming that an aerosol particle consists of *N* monomers of radius $r_{mon}$, the collision rate (*C*) is equal to:

$$C = N\pi r_{mon}^2\, v_{H_2}\, n_{part}, \qquad (14)$$

where $v_{H_2}$ is the mean collision speed of a $H_2$ molecule and $n_{part}$ is the aerosol particle number density. Following Tomasko et al. (2008b) and Doose et al. (2016), we used *N* = 3000 and $r_{mon}$ = 0.05 µm, based on Huygens/DISR measurements. The particle number density ($n_{part}$) follows the vertical extinction profile derived by Doose et al. scaled to a particle number density of 5 cm$^{-3}$ below 55 km (Tomasko et al. 2008b, Doose et al. 2016).

According to Fukutani and Sugimoto (2013), the sticking efficiency *S* of light molecules like hydrogen on a substrate with heavier elements is typically ~0.1. On the other hand, Matar et al. (2010) reported measurements of *S* for $H_2$ on amorphous water ice and found *S* ~ 0.52 at 70-100 K decreasing to 0.34 at 180 K. We used *S* = 0.2 as a nominal case and retain the range 0.1-0.5 for our uncertainty study.

The desorption rate $k_{des}$ is expressed as:

$$k_{des} = \frac{1}{\pi} \sqrt{\frac{2kT_{phys}}{d_0^2\, m_{H_2}}}\, e^{-\frac{T_{phys}}{T}}, \qquad (15)$$

where $T_{phys}$ is the physisorption energy expressed in temperature, $m_{H_2}$ the mass of the $H_2$ molecule, and $d_0$ the width of the potential well, also $\approx n_s^{-1/2}$, that we take equal to 0.3 nm. The physisorption energy depends on the chemical nature of the surface. Measurements have been made for silicates, amorphous carbon and water ice (see discussion in Bron et al. 2016), but not on organic materials which would be more relevant to Titan's aerosols. Following Bron et al. we consider that $T_{phys}$ = 550 K is a reasonable value for both amorphous carbon and amorphous silicates and we take it as our nominal value in this analysis. We also explore the range 300-800 K in our sensitivity study.

Several experiments have been conducted to measure the o-p conversion time of $H_2$ molecules physisorbed on various surfaces: diamagnetic metals, graphite, water ice, with in some cases adsorbed paramagnetic ($O_2$) impurities (see e.g., Section 5 of Fukutani and Sugimoto 2013 and Table 1 of Bron et al. 2016). Although some of the measurements are contradictory, timescales between $10^2$ and $10^3$ s are typical of conversion on amorphous ice and graphite and more generally diamagnetic surfaces. On surfaces with adsorbed impurities, like $O_2$, the conversion time near a paramagnetic site is much shorter, ~10 s (e.g. Fukutani and Sugimoto 2013). Titan's aerosols are made of organic material. Analyses of laboratory analogues of these aerosols show polymeric structures and various functional groups such as nitrile, amine, aliphatic and aromatic hydrocarbons (e.g. Morrison et al. 2016, Sciamma-O'Brien et al. 2017, Maillard et al. 2018). This material should be essentially diamagnetic but it likely contains a small fraction of magnetic sites. These could be radicals present in Titan's atmosphere, like H or $CH_3$, and adsorbed on the surface, or paramagnetic centers created from bond breaks due to solar UV irradiation. Given these uncertainties, we retain $\tau_c = 10^2$ s as a nominal value and consider the whole range $10-10^3$ s as a possibility.

The density of sites has been estimated from various laboratory experiments and vary from $2 \times 10^{14}$ to $1.5 \times 10^{15}$ cm$^{-2}$ (e.g. Biham et al. 2001, Hasegawa et al. 1992, Perets et al. 2007). We used the intermediate value of $6 \times 10^{14}$ cm$^{-2}$ as a nominal case and allowed variation between the two extreme values in our model. As a matter of fact, we found that it is not a critical parameter in Titan's conditions.

The vertical profile of $\tau_{aer}$ is shown in Fig. 11.

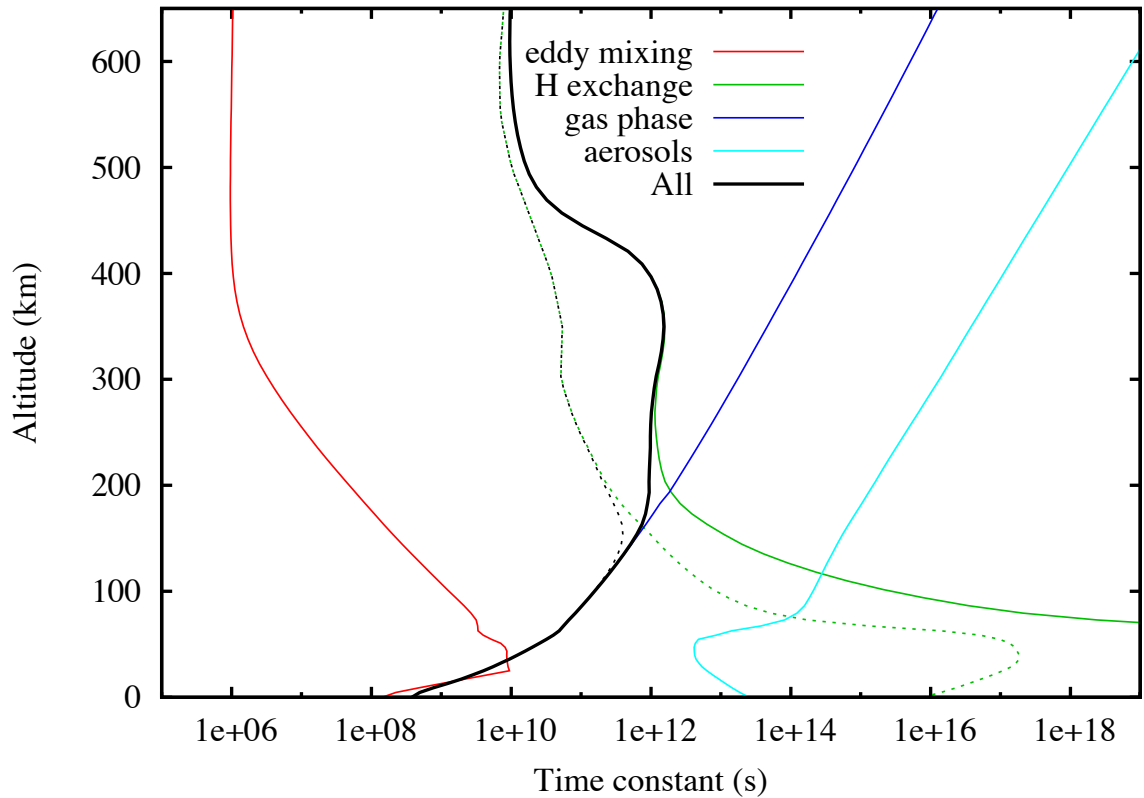

**Fig. 11.** Ortho-para $H_2$ conversion times due to H exchange in the gas phase (green lines), magnetic interaction with $N_2$ and $CH_4$ in the atmosphere (blue line), magnetic interaction on aerosol surface (cyan line), and all these processes together (black lines). The green and black solid lines correspond to the H density profile of Vuitton et al. (2018) and the dashed ones to that of Lavvas et al. (2008). Also shown is the dynamical mixing time calculated from the eddy mixing profile used in our model (red line).

*3.1.4. Net conversion time*

Fig. 11 shows the time constants associated with the three mechanisms we discussed. The net conversion time $\tau_{conv}$, shown in black, is given by:

$$\frac{1}{\tau_{conv}} = \frac{1}{\tau_{exch}} + \frac{1}{\tau_{gas}} + \frac{1}{\tau_{aer}} \qquad (16)$$

H exchange in the atmosphere is the dominant mechanism for the ortho-para $H_2$ conversion in the stratosphere and mesosphere above ~200 km while magnetic interaction with methane, and to a lesser extent dinitrogen, dominates in the lower stratosphere and troposphere. In the troposphere, the conversion time varies from $4 \times 10^8$ s near the surface to $2 \times 10^{10}$ s at 50 km and is similar to the dynamical time within a factor of 3. Ortho-para conversion on the surface of aerosols is not significant. This is mainly because the typical time ($1/k_{des}$) during which a $H_2$ molecule stays in a physisorption state at the surface is much smaller than the conversion time on the surface ($\tau_c$). Even if $\tau_c$ is decreased from $10^2$ to 10 s, the time constant for this process remains at least an order of magnitude larger than that due to interaction of $H_2$ with $CH_4$ and $N_2$. Conversion times associated with exchange of H atoms are larger than the dynamical mixing time by at least 4 orders of magnitude. This is still the case if we use the H profile from Lavvas et al. (2008) in place of our nominal profile from the most recent photochemical model of Vuitton et al. (2019).

*3.2 Model of $H_2$ para fraction profile in Titan's atmosphere*

In this section, we present a one-dimensional model of the $H_2$ para fraction ($f_{para}$) profile in Titan's atmosphere in response to the aforementioned o-p conversion mechanisms. Note that our goal is not to reinvestigate the detailed photochemistry of $H_2$ as in models of e.g. Lavvas et al. (2008), Krasnopolsky (2009, 2014), and Vuitton et al. (2019). We limit our atmospheric grid to 0-500 km, where molecular diffusion can be neglected with respect to eddy diffusion.

We assume that production and loss of H$_2$ takes place only above this altitude range and that the net H$_2$ production rate is equal to its escape rate $E$, so that the net flux of H$_2$ at the upper boundary of our model (500 km) is zero. We also assume no source or sink of H$_2$ at the surface, implying a zero flux at 0 km. The H$_2$ diffusive flux is thus zero and the H$_2$ mole fraction ($q_{H_2}$) constant at all levels of our grid. We fix $q_{H_2}$ to $9 \times 10^{-4}$ as derived from our analysis of CIRS spectra, a value that agrees with the model results of Krasnopolsky (2014) and Vuitton et al. (2019). In addition, the latter model does indicate that ~87% of the H$_2$ production occurs above 500 km and that the H$_2$ mole fraction is essentially constant below this altitude, with a relative variation as low as 1.5% from 0 to 500 km.

The continuity equation for the para-H$_2$ mole fraction $q_{para}$ is

$$-\frac{1}{r^2}\frac{\partial}{\partial z}\left(r^2 K n \frac{\partial q_{para}}{\partial z}\right) = -n\frac{\left(q_{para}-q_{para}^{eq}\right)}{\tau_{conv}}, \qquad (17a)$$

where $z$ is the altitude above the surface of radius $R$, $r$ is equal to $R + z$, $K$ is the eddy mixing coefficient, $n$ is the atmospheric number density, $\tau_{conv}$ the net conversion time in Eq. 16, and $q_{para}^{eq}$ the equilibrium para-H$_2$ mole fraction. The left-hand side of Eq. 17a is the vertical flux divergence while the right-hand side represents the net production rate of para-H$_2$, occurring from o-p conversion. The H$_2$ para fraction being defined as $f_{para} = \frac{q_{para}}{q_{H_2}}$ and $q_{H_2}$ being constant, Eq. (17a) can be written as:

$$-\frac{1}{r^2}\frac{\partial}{\partial z}\left(r^2 K n \frac{\partial f_{para}}{\partial z}\right) = -n\frac{\left(f_{para}-f_{para}^{eq}\right)}{\tau_{conv}}, \qquad (17b)$$

where $f_{para}^{eq}$ is the equilibrium para fraction. We calculate $f_{para}^{eq}$ from Eq. (1) using the rotational constants of H$_2$ in the ground state derived by Jennings et al. (1987). As a lower boundary condition, we assume zero flux of para (and ortho) H$_2$ at the surface, i.e. $-K n \frac{\partial q_{para}}{\partial z} = 0$, which yields $\left.\frac{\partial f_{para}}{\partial z}\right|_{z=0} = 0$. As mentioned above, we consider that the net

H₂ production rate is equal to its escape rate $E$ that we take equal to $1.1 \times 10^{10}$ cm$^{-2}$ s$^{-1}$ (referred to Titan's surface) (Cui et al. 2009). We consider that H₂ is formed by photochemistry with a para fraction at thermodynamical equilibrium at 160 K, $f_{para}^{eq\_prod} = 0.28$. Our upper boundary condition is then that the upward flux of para H₂ is equal to the escape flux of para H₂ minus its production flux. This yields $-K\,n\,\frac{\partial q_{para}}{\partial z} = E\left(f_{para} - f_{para}^{eq\_prod}\right)\left(\frac{R}{r}\right)^2$, or equivalently:

$$\left.\frac{\partial f_{para}}{\partial z}\right|_{z=500\,km} = -\left.\frac{E\left(f_{para}-f_{para}^{eq\_prod}\right)}{q_{H_2}\,K\,n}\left(\frac{R}{r}\right)^2\right|_{z=500\,km} \tag{18}$$

Constraints on the eddy mixing coefficient ($K$) can be obtained from the retrieved vertical profiles of photochemical species using one-dimensional photochemical models (e.g. Vuitton et al. 2019). They indicate that $K$ is small in the lower stratosphere (a few 10³ cm² s⁻¹ at 100 km) and rapidly increases with height up to ~400 km (Vinatier et al. 2007, Lavvas et al. 2008, Vuitton et al. 2019). These observational constraints however only apply down do the condensation levels in the lower stratosphere (~80 km). Dynamical considerations as well as simulations from General Circulation Models indicate that the longest dynamical constants, a few Titan years (equivalent to $K \sim 10^3$ cm² s⁻¹), are found in the lower stratosphere and upper troposphere (Lebonnois et al. 2014). A convective planetary boundary layer (PBL) extends over the first 2 km above the surface, with strong circulation and vertical mixing powered by a sensible heat flux of ~0.2 W m⁻² (Charnay and Lebonnois 2012). Above this region and up to ~25 km, moist convective activity may induce an atmospheric mixing with a strength intermediate between those in the PBL and in the most stable upper troposphere and lower stratosphere (Griffith et al. 2014). Upon these considerations, our nominal $K$ profile is:

Above 45 km: $K = \frac{3\times 10^7}{1\,+\,3.16\times 10^6 p^{1.5}}$ cm² s⁻¹ ($p$ in bar) (Vuitton et al. 2019),

25-45 km: $K = K_{min} = 3 \times 10^2$ cm$^2$ s$^{-1}$,

2-25 km: $K = K_{min}^{(z-2)/23} K_{PBL}^{(25-z)/23}$,

0-2 km: $K = K_{PBL} = 3 \times 10^4$ cm$^2$ s$^{-1}$.

Note that our $K_{PBL}$ is about one order of magnitude lower that the value for Earth's PBL.

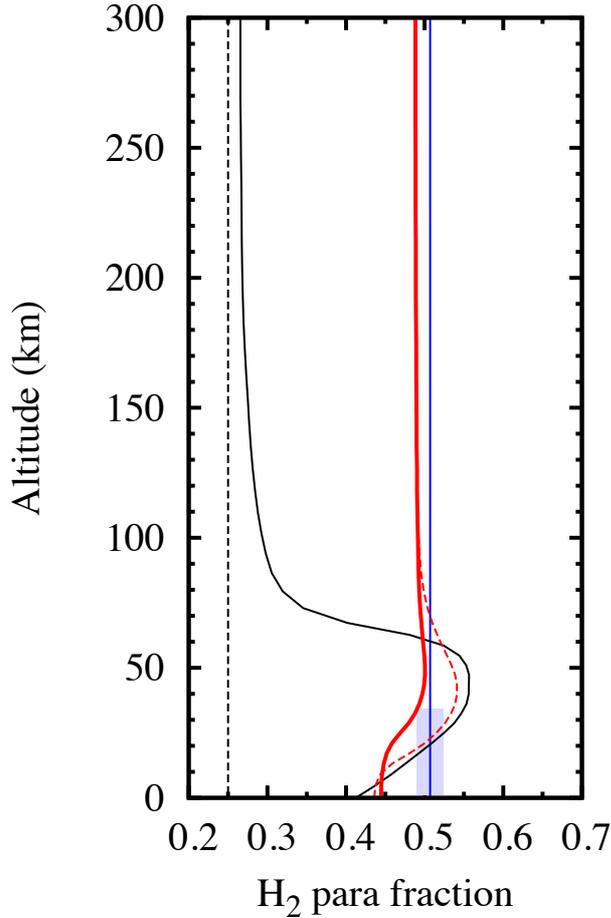

**Fig. 12.** H$_2$ para fraction ($f_{para}$) as a function of altitude. <u>Black solid line:</u> o-p local thermodynamical equilibrium; <u>Thick red line:</u> nominal o-p equilibration model; <u>Dashed red line:</u> "adjusted" o-p equilibration model (see text). The blue line represents the constant para H$_2$ profile that fits the observations as well as the o-p equilibrium profile. The blue box represents the 1-SD error bars about this constant profile and the vertical range of sensitivity of the measurements. The black dashed line shows the high-temperature limit of the H$_2$ para fraction (0.25).

Fig. 12 shows the vertical profile of the H$_2$ para fraction calculated with this nominal model (thick red line). The profile shows relatively little variation in the troposphere, increasing from 0.44 at the surface to 0.50 at 50 km. Level-by-level equilibration is not realized because the conversion time is not short enough compared with the vertical mixing time. In the stratosphere, above 70 km, the H$_2$ para fraction is ~0.49, which is 75% above its equilibrium value of ~0.28. As the conversion time is much larger than the dynamical mixing time, the H$_2$ para fraction is not representative of the local equilibrium value but reflects a value characteristic of tropospheric levels where conversion occurs on a timescale similar to the mixing one.

We made a first test in which we decreased the escape rate $E$ in Eq. (18) from $1.1 \times 10^{10}$ cm$^{-2}$ s$^{-1}$ to $0.57 \times 10^{10}$ cm$^{-2}$ s$^{-1}$, a value that corresponds to Jeans escape in Vuitton et al.'s (2019) photochemical calculations (V. Vuitton, private communication). In this case, we found a maximum relative decrease of $f_{para}$ of only 1% above 100 km. We then tested the sensitivity of our H$_2$ para fraction profile to various parameters (Table 2). The reaction rates in Eqs. 9-10 for o-p conversion with N$_2$ and CH$_4$ have not been measured in the laboratory and have been extrapolated here from measurements of the reaction rate with ortho H$_2$. We then multiplied the time constant $\tau_{gas}$ by factors of 5 and 1/5, leaving all other parameters unchanged, and found a maximum change in the latter case of $f_{para}$ from 0.46 to 0.53 at 50 km and from 0.46 to 0.49 in the stratosphere. In a second test, we replaced the H profile of Vuitton et al. (2019) by that of Lavvas et al. (2008) which exhibits some 20 times larger densities in the upper stratosphere and lower mesosphere. This change does not alter the H$_2$ para fraction by > 0.01 at any level. The uncertainties in the parameters that govern hydrogen o-p conversion on Titan's aerosols have negligible effect on the model results. A test, in which we used extreme values, $S = 0.5$, $\tau_c = 10$ s, $T_{phys} = 800$ K to maximize the conversion efficiency and minimize $\tau_{aer}$, does not show any increase of $f_{para}$ in excess of 0.01 at any level (Col. 7 of Table 2). Finally we made tests in which we changed the values of the eddy mixing coefficients around the tropopause $K_{min}$ and near the surface $K_{PBL}$ by factors of 3 and 1/3 (Cols. 8-11 of Table 2). Again the effects are weak with an increase of $f_{para}$ by ~0.02 at the tropopause and above, when a low value of $K_{min}$ ($10^2$ cm$^2$ s$^{-1}$) is used and a decrease by ~0.03 when a high value of $K_{min}$ ($10^3$ cm$^2$ s$^{-1}$) is used. In all cases tested, the H$_2$ para mole fraction below 20 km is almost unchanged. At the tropopause, we get a range of 0.47-0.52 while, between 100 and 400 km, the range of variation is 0.46-0.51.

Our nominal model $H_2$ para profile, along with a $H_2$ mole fraction of to $0.88 \times 10^{-3}$, allows us to fit the CIRS FP1 spectra only at the 4-SD error level. In this nominal model, $f_{para}$ is ~10% too low in the mid-troposphere to reproduce the CIRS results. In fact, for all the cases that we tested, the calculated $H_2$ para fraction does not follow closely the equilibrium profile even in the troposphere. In the first 35 km, it is also below the range 0.49-0.53 needed to reproduce the observations using a constant-with-height profile. To reach a value in this range at 20 km, and accordingly a profile that would agree with the CIRS spectra, it is necessary to use model parameters that deviate from their nominal values further than we assumed. For example, dividing $\tau_{gas}$ by a factor of 20 would produce $f_{para}$ = 0.49 at 20 km. Another possibility is to strongly reduce the mixing strength in the 20-km region. We have then designed an "adjusted" case in which we extended the region where $K = K_{min} = 3 \times 10^2$ cm$^2$ s$^{-1}$ down to 15 km (instead of 25 km) and divided $\tau_{gas}$ by a factor of 5. In this "adjusted" case (last column of Table 2 and dashed red line in Fig. 12), the $H_2$ para profile follows more closely the equilibrium profile in the troposphere and has $f_{para}$ = 0.49 at 20 km, which allows us to fit the CIRS FP1 spectra at the 1.4-SD precision level.

**Table 2.**

$H_2$ para fraction at different altitudes according to various model parameters

| Altitude (km) | Equi-librium value | Nominal model | $\tau_{gas}$ × 5 | $\tau_{gas}$ / 5 | H from Lavvas et al. | $\tau_{aer}$ min | $K_{min}$ = $1 \times 10^2$ cm$^2$ s$^{-1}$ | $K_{min}$ = $1 \times 10^3$ cm$^2$ s$^{-1}$ | $K_{PBL}$ = $10^4$ cm$^2$ s$^{-1}$ | $K_{PBL}$ = $10^5$ cm$^2$ s$^{-1}$ | "Adjusted" |
|---|---|---|---|---|---|---|---|---|---|---|---|
| 0 | 0.41 | 0.44 | 0.45 | 0.44 | 0.44 | 0.44 | 0.44 | 0.45 | 0.44 | 0.44 | 0.44 |
| 10 | 0.46 | 0.45 | 0.45 | 0.45 | 0.45 | 0.45 | 0.45 | 0.45 | 0.45 | 0.45 | 0.45 |
| 20  | 0.50 | 0.46 | 0.45 | 0.47 | 0.46 | 0.46 | 0.46 | 0.45 | 0.46 | 0.45 | 0.49 |
| 50  | 0.55 | 0.50 | 0.46 | 0.53 | 0.50 | 0.50 | 0.52 | 0.47 | 0.50 | 0.50 | 0.53 |
| 100 | 0.29 | 0.49 | 0.46 | 0.49 | 0.49 | 0.49 | 0.51 | 0.47 | 0.49 | 0.49 | 0.49 |
| 200 | 0.27 | 0.49 | 0.46 | 0.49 | 0.49 | 0.49 | 0.51 | 0.46 | 0.49 | 0.49 | 0.49 |
| 400 | 0.27 | 0.49 | 0.46 | 0.49 | 0.49 | 0.49 | 0.50 | 0.46 | 0.49 | 0.49 | 0.49 |

## 4. Discussion

*4.1. Collision-induced absorption*

Having constrained the opacity profiles of $CH_3C_2H$, $C_4H_2$, the main haze and the nitrile haze from CIRS FP1 and FP3 limb spectra, we found that fitting of the nadir FP1 spectra required some adjustments of the $N_2$-$CH_4$ and $N_2$-$N_2$ CIA with respect to the quantum mechanical calculations of Borysow and Tang (1993) and Borysow and Frommhold (1986a) respectively.

We derived that a multiplicative factor of 1.52, with a typical uncertainty of ± 0.05, was needed for the $N_2$-$CH_4$ coefficients at low temperature (70-85 K) to reproduce the observed continuum below 325 cm$^{-1}$. This conclusion fully agrees with the conclusions of Tomasko et al. (2008a) and de Kok et al. (2010) who also suggested a factor of ~1.5 correction. On the other hand, the wavenumber-dependent factor proposed by Anderson and Samuelson (2011) produces too much absorption between 180 and 300 cm$^{-1}$ (Fig. 6) and more generally a spectral shape that does not match the observations. Our result is actually also consistent with the analysis of Samuelson et al. (1997) of Voyager/IRIS spectra who argued for a methane supersaturation of 1.53 ± 0.09 in the upper troposphere. Since they used the CIA model from Borysow and Tang (1993), including a $CH_4$ supersaturation factor of 1.5 is equivalent to using a saturated CH$_4$ profile (as in the Niemann et al. 2010 profile) combined with 1.5-time larger CIA coefficients as we did. Anderson and Samuelson (2011) argue that the wavenumber dependence of the correction factor they propose ($\beta_\nu$) is similar to that of the extra optical thickness ($\delta\tau_\nu$) needed by Samuelson et al. (1997) when CH$_4$ supersaturation is not allowed. However, the correct approach is to compare the spectral variation of $\delta\tau_\nu$ with $\beta_\nu \times k_\nu$, where $k_\nu$ is an average N$_2$-CH$_4$ CIA coefficient at tropospheric temperatures. According to Samuelson et al., $\delta\tau_\nu$ has the characteristics of methane absorption, i.e. of $k_\nu$, so that this comparison points again to an essentially constant value of $\beta_\nu$.

As concerns N$_2$-N$_2$, we showed that the use of a corrective factor to Borysow and Frommhold's (1986a) beyond 110 cm$^{-1}$ (Eq. 3) to agree with the quantum mechanical calculation of Karman et al. (2015) better reproduces the FP1 spectra in the range ~140-270 cm$^{-1}$. The refined quantum mechanical calculations of Karman et al. (2015) based on the Close-Coupling method and including the full anisotropic potential are clearly superior to those of Borysow and Frommhold (1986a) who used effective isotropic potentials and adjusted collision-induced dipole moments (see review by Hartmann et al. 2018). We therefore recommend to use the corrective factor in Eq. 3 to model Titan's spectrum (or directly the coefficients in Table III of Karman et al. beyond 110 cm$^{-1}$). Note that laboratory measurements relevant to Titan's tropospheric temperatures have been performed at 78 and 93 K (Wishnow et al. 1996) but are limited to wavenumbers < 100-125 cm$^{-1}$, which does not permit to confirm yet Karman et al.'s calculations in the N$_2$ band far wing.

Our best fit model still shows some disagreement with the observed spectra, especially around 120-150 cm-1 where our model overestimates the observed emission at both high and low emission angle by ~2%. It is difficult to assess the origin of this little misadjustment. It could be due to remaining errors in the $N_2$-$N_2$ and/or $CH_4$-$N_2$ CIA coefficients combined with a slightly too large nitrile haze opacity. On the other hand, we can match both FP1 spectra below 100 cm$^{-1}$ using the nominal $N_2$-$N_2$ coefficients, in contrast with Anderson and Samuelson (2011) who invoke either the need for large reductions of these coefficients below 80 cm$^{-1}$ or faulty instrument calibration.

*4.2. $H_2$ mole fraction*

The $N_2$-$H_2$ collision-induced $S_0(0)$ and $S_0(1)$ $H_2$ lines in the FP1 nadir spectra can be fitted equally well (at the 1-SD error level) assuming o-p equilibrium or a constant $H_2$ para fraction. The derived $H_2$ mole fractions in both cases agree within 1-SD error bars, $(0.875 \pm 0.03) \times 10^{-3}$ and $(0.89 \pm 0.03) \times 10^{-3}$, and we thus conclude that it lies between $0.84$-$0.92 \times 10^{-3}$, accounting only for measurement errors. This value pertains to tropospheric levels below ~35 km. It is fully consistent with the previous determination from CIRS spectra by Courtin et al. (2012), $(0.96 \pm 0.24) \times 10^{-3}$, based on the $H_2$-$N_2$ dimer rotational transitions around 355 cm$^{-1}$. The $H_2$ mole fraction derived by Samuelson et al. (1997) from Voyager IRIS data using the Borysow et al. (1986b) CIA coefficients and assuming methane supersaturation is $(1.12 \pm 0.16) \times 10^{-3}$. This is only 1.5 SD above our determination, which we regard as a satisfactory agreement. A preceding analysis of Voyager IRIS spectra by Courtin et al. (1995) indicated a $H_2$ mole fraction of $(1.3 \pm 0.3) \times 10^{-3}$ using only the $S_0(0)$ line and assuming methane supersaturation, which is again only 1.5 SD above the present determination.

The $H_2$ abundance we derived bears an uncertainty due to possible inaccuracies in the $H_2$-$N_2$ CIA coefficients. The latter were estimated to be ± 10-15% by Samuelson et al. (1997), based on the agreement between the laboratory measurements of Dore et al. (1986) from 91 to 298 K and the theoretical model of Borysow et al. (1986b). Dore et al. mention that "the agreement is typically better than the uncertainty of the measurement (± 10%; and ± 15% at 90 K)", which suggests that the actual uncertainty of the model and of the measurement might be smaller than quoted. On the other hand, Dore et al. note that, on the average, theoretical intensities near the peaks of the $S_0(0)$ and $S_0(1)$ lines are "slightly excessive", which they attribute to a small anisotropic overlap component not included in the Borysow et al. model. In fact, at 91 K, the model exceeds the observations at the peaks of the $S_0(0)$ and $S_0(1)$ lines by ~15%, which is similar to the quoted uncertainty of the measurement. If the peak absorption coefficients we use in our model are indeed 15% too large, then the $H_2$ mole fraction we derived should be increased by this amount, i.e. up to ~$1.0 \times 10^{-3}$. In the lack of more precise measurements, we simply assume a ± 15% uncertainty on the $H_2$ mole fraction, which leads to $(0.88 \pm 0.13) \times 10^{-3}$.

Our $H_2$ determination agrees with the in situ measurement by the Huygens/GCMS, which yields a uniform mole fraction of $(1.01 \pm 0.16) \times 10^{-3}$ below 140 km (Niemann et al. 2010). Our study thus confirms that the $H_2$ mole fraction in the troposphere is 3 to 4 times smaller than inferred from in situ measurements by Cassini/INMS from 981 to 1151 km (Cui et al. 2009). From known photochemistry and in the lack of a significant surface sink of $H_2$, such a large difference cannot be maintained given the characteristics of molecular and thermal diffusion in Titan's atmosphere, and assuming "standard" eddy mixing profile. According to Strobel's (2010, 2012) modelling, the tropospheric (CIRS) and thermospheric (INMS) values are incompatible by a factor of 2-3. If the INMS $H_2$ measurements are correct, this discrepancy points to some unidentified physical or chemical mechanism at work to deplete $H_2$ in the lower atmosphere or a peculiar choice of the eddy mixing profile as proposed by Krasnopolsky (2012, 2014).

*4.3. $H_2$ ortho-to-para ratio*

This study provides the first precise constraint on the hydrogen para fraction in Titan's troposphere using Cassini measurements. We conclude that the o-p profile in the lowest 35 km is fully consistent with equilibrium within error bars. On the other hand, the spectra can be almost equally well fitted with a uniform para fraction equal to 0.51 ± 0.02. The allowed range is representative of o-p equilibrium at temperatures between 74 and 81 K, which are reached at 12-25 km. This altitude range essentially corresponds to the peak of sensitivity for the $S_0(0)$ line (Fig. 10). Although the altitudes probed by the $S_0(0)$ and $S_0(1)$ lines differ by some 10 km, we were not able to distinguish between a uniform $H_2$ para fraction of ~ 0.51 and the altitude-dependent equilibrium profile. This is due to the measurement errors and remaining uncertainties in the radiative transfer modelling.

In their analysis of the $S_0(0)$ and $S_0(1)$ $H_2$ lines from Voyager/IRIS spectra, Courtin et al. (1995) concluded that there was no evidence for a departure from equilibrium and that the para fraction was most likely larger than the high-temperature limit (0.25). Still from Voyager/IRIS spectra, Samuelson et al. (1997) inferred that the $H_2$ para fraction was close to equilibrium, with considerable uncertainty. In fact, their best fit (assuming methane supersaturation) is obtained for a $H_2$ para fraction in the range 0.36-0.51 (1-SD confidence level), which is consistent with our finding.

*4.4. $H_2$ ortho-para equilibration*

We have investigated different processes that can lead to the equilibration of the ortho-to-para ratio in $H_2$ in Titan's atmosphere: H exchange in the gas phase, o-p conversion on the aerosols, magnetic interaction with $N_2$ and $CH_4$. We found that, while H exchange is the dominant process above 160-200 km, it is slower than the atmospheric mixing time by 3 to 4 orders of magnitude and thus cannot ensure o-p equilibrium at these altitudes. The vertical profile of H atoms we considered derives from photochemical models and is thus relatively uncertain: for example, those from Lavvas et al. (2008) and Vuitton et al. (2019) differ by an order of magnitude below 400 km. However, the increase in H atom density that would be required to bring the conversion time near the vertical mixing time is unrealistically large.

At lower levels, the dominant process is magnetic interaction of $H_2$ with $CH_4$ and to a lesser extent $N_2$. In the troposphere, the time constant associated with this process is similar to the estimated eddy mixing time, which allows at least partial equilibration of the o-p ratio. Conversion on aerosol surface reaches its maximum efficiency around the tropopause, where temperature is minimum and thus the desorption rate lowest. However, the associated time constant in this region is still 3 orders of magnitude larger than the dynamical mixing time in our nominal model.

We have developed a 1-D model of the $H_2$ para fraction, which solves the continuity equation in the presence of the various conversion mechanisms. This allowed us to investigate the effects of key uncertain model parameters in the conversion processes themselves and in the eddy mixing properties. The efficiency of ortho-para hydrogen conversion on Titan's aerosols depends on parameters that are relatively uncertain: sticking efficiency, physisorption energy and conversion time constant in a physisorbed state. We found that in all cases where we varied these parameters in a reasonable range, the $H_2$ para fraction profile was unaffected by this conversion process. This is still the case if we decrease the conversion time to 1 s, which could be relevant if the aerosol surface includes a significant fraction of paramagnetic sites.

In our model, the $H_2$ para fraction is governed by H exchange with $CH_4$ and $N_2$ in the troposphere. The conversion efficiency depends on both the reaction rates (R1a and R1b) and the strength of the atmospheric vertical mixing. In our nominal case, the $H_2$ para fraction increases from 0.44 at the surface to 0.50 at the tropopause, but is still ~10% too low to reproduce the Titan observations. The reaction rates have not been measured in the laboratory and we had to extrapolate them from those for $H_2$ (Milenko et al. 1997) to $CH_4$ and $N_2$, based on the change in magnetic moment and Lennard-Jones potential parameters. This estimation bears significant uncertainty. However, reproducing the CIRS results would require to use 20 times more rapid reactions than in our estimate (leaving other parameters unchanged), which seems a large change. The atmospheric mixing below 50 km is poorly constrained. Decreasing our minimal value of the eddy mixing coefficient $K_{min}$ by a factor of 3 in the range 25-45 km, down to 100 cm$^2$ s$^{-1}$, does not increase significantly the $H_2$ para fraction below 35 km where the $H_2$ $S_0(0)$ and $S_0(1)$ lines are formed. Similarly, decreasing $K_{PBL}$ at the surface by a factor of 3, down to $10^4$ cm$^2$ s$^{-1}$, does not significantly affect the para-fraction profile. On the other hand, extending the region where $K = K_{min}$ down to ~15 km or below has a more pronounced effect. An "adjusted" case that we have designed, with $K_{min} = 300$ cm$^2$ s$^{-1}$ down to 15 km and reactions rates for R1 five times larger than we estimated, yields a para fraction profile that allows us to reproduce the CIRS results within error bars.

Therefore, we believe that the CIRS results, which indicate equilibration of o-p $H_2$ at least in the 20-km region, point to both slow vertical mixing in the troposphere down to at least ~15 km and to an equilibration rate with $CH_4$ and $N_2$ slightly higher that our estimation.

## 5. Conclusions

We have analyzed two selections of CIRS spectra between 50 and 650 cm$^{-1}$, at low and high emission angles, to determine both the H$_2$ abundance and its ortho-to-para ratio in Titan's troposphere. This was done by simultaneously considering the absorption features around the $S_0(0)$ line near 355 cm$^{-1}$, due to para-H$_2$ and the $S_0(1)$ line, due to ortho-H$_2$. To better constrain the profiles of CH$_3$C$_2$H, C$_4$H$_2$, the main haze and the nitrile haze, which are important contributors to the opacity around these H$_2$ lines, we also used CIRS FP1 and FP3 limb spectra. From these limb spectra, we derived the spectral dependence of the main haze extinction (Fig. 4). Below 610 cm$^{-1}$, we recommend to use our (relative) extinction coefficients in place of those of Vinatier et al. (2012) which are more affected by noise measurements.

Following Tomasko et al. (2008a) and de Kok et al. (2010), we confirm that the N$_2$-CH$_4$ collision-induced absorption predicted by Borysow and Tang (1993) is too weak at temperatures of 70-85 K. We found that a constant multiplicative factor of ~1.52 yields a satisfactory reproduction of the CIRS spectra, in contrast with the conclusions of Anderson and Samuelson (2011) which were essentially based on limb and high emission angle spectra. Existing laboratory measurements of N$_2$-CH$_4$ CIA are limited to temperatures above 126 K (Dagg et al. 1986) and there is thus a strong need for new measurements down to 70 K to settle the issue. In complement, ab initio quantum mechanical calculations should be performed to predict the N$_2$-CH$_4$ absorption spectrum as a function of temperature for use in radiative transfer calculations.

We also think that the $N_2$-$N_2$ collision-induced absorption calculated from Borysow and Frommhold (1986a) is too weak in the far wing of the $N_2$ band, beyond 110 cm$^{-1}$. We propose a corrective factor (Eq. 2) to agree with the quantum mechanical calculation of Karman et al. (2015), which are free of empirically adjusted parameters. Existing $N_2$-$N_2$ laboratory data at temperatures pertaining to Titan's troposphere do not extend beyond 125 cm$^{-1}$ (Wishnow et al. 1996) and measurements up to ~250-300 cm$^{-1}$ would thus be needed to confirm our conclusion.

With these adjustments, we were able to reproduce quite well both the low and high emission angle CIRS spectra in the whole FP1 range above 50 cm$^{-1}$. The remaining discrepancy, of the order of 2% from 120 to 150 cm$^{-1}$, could stem from remaining errors in the CIA coefficients and in the nitrile haze opacity profile.

Our best fitting $H_2$ mole fraction is $(0.88 \pm 0.13) \times 10^{-3}$, including a ± 15% uncertainty on the $N_2$- $H_2$ CIA coefficients near the peaks of the $S_0(0)$ and $S_0(1)$ lines. This value pertains to the troposphere, mostly between 1 and 34 km (Fig. 10). This value agrees with the previous determination from CIRS spectra by Courtin et al. (2012), which was solely based on the $H_2$-$N_2$ dimer $S_0(0)$ rotational transition around 355 cm$^{-1}$. It is also consistent with older determinations from Voyager infrared spectra that assumed $CH_4$ supersaturation along with unmodified $N_2$-$CH_4$ CIA from Borysow and Tang (1993). We therefore confirm that the $H_2$ mole fraction in the troposphere is 3-4 times smaller than the Cassini/INMS determination at 1000-1100 km (Cui et al. 2009). Considering known photochemical and diffusion processes, Strobel (2010, 2012) concluded that the two measurements are incompatible by a factor of 2-3. If the INMS result is correct, this difference points to a sink for $H_2$ in the lower atmosphere or at the surface that remains to be identified.

Our analysis indicates that the ortho-to-para ratio of $H_2$ is close to equilibrium in the 20-km region. The CIRS spectra can be fitted almost equally well assuming either an equilibrium o-p profile in the whole atmosphere or a uniform para fraction between 0.49 and 0.53. This interval corresponds to the equilibrium value at altitudes in the range 12-25 km, where our $H_2$ retrievals are most sensitive. The present analysis is the first to derive relatively stringent constraints on the o-p ratio in Titan's atmosphere.

From a careful survey of the o-p conversion mechanisms potentially at work in Titan's atmosphere, we conclude that magnetic interaction with $CH_4$, and to a lesser extent $N_2$, is the only one that can operate in the troposphere on a timescale comparable with that of dynamical mixing. H exchange and conversion through adsorption and magnetic interaction on aerosol surface are not efficient enough in regard to atmospheric mixing. We have developed a 1-D model for the $H_2$ para fraction that solves the continuity equation and incorporates the conversion mechanisms in terms of time constants. Our nominal model, based on our best estimates for the eddy mixing profile and conversion reaction rates, produces a para fraction profile that is about 10% smaller than required to fit the observations at the 1-SD level. In fact, to reach this level of agreement, the model needs a low eddy mixing coefficient (~300 cm$^2$ s$^{-1}$ or less) from the tropopause down to at least ~15 km and equilibration rates from interaction of $H_2$ with $CH_4$ and/or $N_2$ somewhat larger than estimated (by a factor of ~5). These rates have not been measured in the laboratory and we had to extrapolate them from experimental data on natural conversion in gaseous hydrogen (Milenko et al. 1997). Laboratory measurements of o-p conversion rates through interaction of $H_2$ with $CH_4$ and $N_2$ would thus be valuable.

We finally note that our model predicts that $H_2$ is far from equilibrium above ~90 km with a para fraction of ~0.49 while the equilibrium value lies between 0.27 and 0.30. This behavior stems from the inefficiency of conversion mechanisms in the stratosphere and mesosphere while equilibration can occur in the mid-troposphere where temperatures are lower. This prediction does not seem however easily testable from remote sensing observations.

**Acknowledgments**

We acknowledge support from the Centre National d'Études Spatiales (CNES) and the Programme National de Planétologie (INSU/CNRS). We are indebted to the CIRS instrument team for planning the observations and calibrating the data. We thank V. Vuitton for providing files of her atmospheric model, H profile and $H_2$ production rate profile, and F. Le Petit for very useful discussions on the ortho-para conversion of $H_2$ on interstellar grain surfaces.